\documentclass[aps,prc,
showpacs,twocolumn,superscriptaddress,nofootinbib]{revtex4-1}

\usepackage{graphicx}
\usepackage{colordvi}
\usepackage{bm}
\usepackage{amsmath}
\usepackage{amssymb}
\usepackage{verbatim}
\usepackage{multirow}
\usepackage{bigstrut}
\usepackage{array}
\usepackage{fix-cm}
\usepackage{bbold}

\newcommand{\mC}{\mathcal{C}}
\newcommand{\mD}{\mathcal{D}}

\newcommand{\mM}{\mathcal{M}}
\newcommand{\mL}{\mathcal{L}}
\newcommand{\mN}{\mathcal{N}}

\newcommand{\mU}{\mathcal{U}}
\newcommand{\mV}{\mathcal{V}}

\newcommand{\mW}{\mathcal{W}}
\newcommand{\mZ}{\mathcal{Z}}

\newcommand{\ii}{\textrm{i}}

\newcommand{\nmax}{N_{\mbox{\footnotesize{max}}}}

\begin{document}

\title{{\it Ab-initio} self-consistent Gorkov-Green's function calculations of semi-magic nuclei \\ II. Numerical implementation at second order with a two-nucleon interaction}

\author{V. Som\`a}
\email{vittorio.soma@physik.tu-darmstadt.de}
\affiliation{Institut f\"ur Kernphysik, Technische Universit\"at Darmstadt, 64289 Darmstadt, Germany}
\affiliation{ExtreMe Matter Institute EMMI, GSI Helmholtzzentrum f\"ur Schwerionenforschung GmbH, 64291 Darmstadt, Germany}

\author{C. Barbieri}
\email{c.barbieri@surrey.ac.uk}
\affiliation{Department of Physics, University of Surrey, Guildford GU2 7XH, UK}

\author{T. Duguet}
\email{thomas.duguet@cea.fr}
\affiliation{CEA-Saclay, IRFU/Service de Physique Nucl\'eaire, 91191 Gif-sur-Yvette, France}
\affiliation{National Superconducting Cyclotron Laboratory and Department of Physics and Astronomy,
Michigan State University, East Lansing, MI 48824, USA}

\begin{abstract}

\begin{description}

\item[Background]
The newly developed self-consistent Gorkov-Green's function approach represents a promising path to the \textit{ab initio} description of mid-mass open-shell nuclei. The formalism based on a two-nucleon interaction and the second order truncation of Gorkov's self-energy has been described in detail in Ref.~[V. Som\`a, T. Duguet and C. Barbieri, Phys.\ Rev.\ C \textbf{84}, 064317 (2011)].
\item[Purpose]
The objective is discuss the methodology used to solve Gorkov's equation numerically and to gauge its performance in view of carrying out systematic calculations of medium-mass nuclei in the future. In doing so, different sources of theoretical error and degrees of self-consistency are investigated. 
\item[Methods]
We employ Krylov projection techniques with a multi-pivot Lanczos algorithm to efficiently handle the growth of poles in the one-body Green's function that arises as a result of solving Gorkov's equation self-consistently. We first characterize the numerical scaling of Gorkov's calculations based on full self-consistency and on a partially self-consistent scheme coined as "sc0". Using small model spaces, the Krylov projection technique is then benchmarked against exact diagonalization of the original Gorkov matrix. Next, the convergence of the results as a function of the number $N_\ell$ of Lanczos iterations per pivot is investigated in large model spaces. Eventually, the convergence of the calculations with the size of the harmonic oscillator model space is examined.
\item[Results]
Gorkov SCGF calculations performed on the basis of Krylov projection techniques display a favourable numerical scaling that authorizes systematic calculations of mid-mass nuclei. The Krylov projection selects efficiently the appropriate degrees of freedom while spanning a very small fraction of the original space. For typical large-scale calculations of mid-mass nuclei, a Krylov projection making use of $N_\ell\approx 50$ yields a sufficient degree of accuracy on the observables of interest. The partially self-consistent sc0 scheme is shown to reproduce fully self-consistent solutions in small model spaces at the 1$\%$ level. Eventually, Gorkov-Green's function calculations performed on the basis of SRG-evolved interactions show a fast convergence as a function of the model-space size.
\item[Conclusions]
The end result is a tractable, accurate and gently scaling \textit{ab initio} scheme applicable to complete isotopic and isotonic chains in the medium-mass region. The partially self-consistent sc0 scheme provides an excellent compromise between accuracy and computational feasibility and will be the workhorse of systematic Gorkov-Green's function calculations in the future.  The numerical scaling and performances of the algorithm employed offers the possibility (i) to apply the method to even heavier systems than those (e.g. $^{74}$Ni) already studied so far and (ii) to perform converged Gorkov SCGF calculations based on harder, e.g. original Chiral, interactions.

\end{description}
\end{abstract}


\maketitle


\section{Introduction}

In the last decade nuclear structure theory has been characterized by remarkable developments in \textit{ab initio} calculations beyond the lightest elements. Approaches like coupled cluster (CC) \cite{Hagen:2012sh, Hagen:2012fb, Binder:2012mk}, Dyson self-consistent Green's functions (SCGF) \cite{Barbieri:2009ej, Barbieri:2012rd, Cipollone:2013b} and in-medium similarity renormalization group (IM-SRG) \cite{Tsukiyama:2010rj, Hergert:2013uja} are nowadays able to successfully describe properties of nuclei in the region $A\sim15\!-\!50$ starting solely from the knowledge of elementary two- and three-nucleon forces.
Such methods, while differing in the way they solve the many-body Schr\"odinger equation, produce results with a similar degree of accuracy, e.g. for ground-state energies in the oxygen chain \cite{Hagen:2012sh, Cipollone:2013b, Hergert:2013uja}. Even more recently, nuclear lattice effective field theory (NLEFT) has joined the group of promising ab-initio methods applicable to mid-mass nuclei~\cite{Lahde:2013uqa}. 

We focus here on SCGF theory whose implementation within Dyson's formalism has been typically limited to doubly closed-shell nuclei so far\footnote{A key feature of SCGF theory is to access the spectral strength distribution associated with one neutron/proton addition or removal, i.e. it automatically delivers the spectrum of  $\text{A}\pm1$ systems out of the calculation of the A-body ground state~\cite{Dickhoff:2004xx, Rios:2009gb, Rios:2011zd}.}. In a few cases superfluid systems have been addressed within the Nambu-Gorkov formalism by including quasiparticle-phonon couplings in the self-energy, either phenomenologically \cite{VdSluys1993} or in the framework of nuclear field theory \cite{Idini2012}. Recently, we have introduced a fully \textit{ab initio} approach based on the Gorkov ansatz that extends the SCGF formalism to open-shell nuclei~\cite{Soma:2011GkvI,Soma:2013rc}. Together with the latest advances on elementary inter-nucleon interactions, such a development paves the way for an \textit{ab initio} description of complete isotopic and isotonic chains in the mid-/heavy-mass region of the nuclear chart.

A crucial issue for \textit{ab initio} approaches concerns the ability of performing numerical calculations in increasingly large model spaces, with the aims of thoroughly checking the convergence and of constantly extending the reach to heavier systems. 
More generally, \textit{ab initio} methods must eventually assess all sources of theoretical uncertainties and attribute theoretical error bands to their predictions. This is a necessary condition to be in the position of exploiting the remaining discrepancy with experiment as a measure of the quality of the input many-body Hamiltonian. The intent of the present work is to discuss the numerical implementation of Gorkov-Green's function techniques for finite systems and evaluate uncertainties associated with model-space truncations and the algorithm used to solve Gorkov's equation. Other sources of error, including uncertainties related to renormalization group transformations of the Hamiltonian and to many-body truncations have already been discussed in the literature~\cite{Cipollone:2013b, Hergert:2013uja} and will be addressed thoroughly for Gorkov theory in future works.

A long-standing problem with self-consistent calculations of one-body propagators in finite systems concerns the rapid increase of the number of poles generated at each iterative step. The fast growth is expected as the Lehmann representation of one-body Green's functions (see Eqs.~\eqref{eq:leh} and \eqref{eq:leh_self} below) develops a continuous cut along the real energy axis in connection with unbound states. 
This cut is discretized by a growing number of discrete energy states as the the size of the model space is increased.  In practical calculations, one needs to limit the number of discretized poles in a way that self-bound systems can still be accurately calculated.
Traditionally, this has been achieved by either binning the self-energy poles along the energy axis or by employing Lanczos algorithms to project the energy denominators onto smaller Krylov spaces~\cite{VanNeck:1991,VanNeck:1993,Muether1993,Muether1995,Dewulf:1997Bagel,VanNeck2001}. 
The latter approach is preferable since the original self-energy is retrieved in the limit of increasing Krylov basis size. However, corresponding calculations  relied on the further approximation that the self-energy is diagonal in the one-body Hilbert space. 
This approximation can result in significant inaccuracies and should be avoided. Moreover, several pivots are necessary to correctly reproduce the off-diagonal features of the self-energy, leading to a block Lanczos algorithm~\cite{Schirmer1989}.
Other works have avoided Krylov projection techniques and performed self-consistent calculations by manually selecting the set of poles carrying the largest strength while collecting the others into few effective poles. These \textit{ad hoc} procedures have led to successful investigations~\cite{Barbieri2002,Barbieri:2006sq} but do not offer the possibility to systematically assess errors.

Our recent SCGF calculations~\cite{Barbieri:2009nx,Waldecker:2011by,Soma:2013rc,Cipollone:2013b} have relied on modified Lanczos and Arnoldi algorithms to perform reduction to Krylov spaces defined by multiple pivots, as originally suggested in Ref.~\cite{Schirmer1989}. This approach guarantees convergence to the full original self-energy in the limit of increasing Krylov space dimension and, hence, is suitable for \textit{ab initio} calculations. However, no account has been given so far of the performance and accuracy of this method in nuclear structure applications. One aim of the present work is to fill this gap.

The paper is organized as follows. In Sec. \ref{sec:GGFtheory} Gorkov-Green's function theory is briefly
reviewed, with a focus on the aspects inherent to the solution of Gorkov's equation.
In Sec. \ref{sec:numerical} the numerical implementation of Gorkov's equation is discussed, 
with particular emphasis on the modified Lanczos
algorithm employed in the diagonalization.
A remainder of the relevant Lanczos formulae as well as details on the treatment of chemical potentials can be found in the Appendix.
The performance of the Krylov projection is analyzed in Sec. \ref{sec:lanczos}.
In Sec. \ref{sec:self} different degrees of self-consistency in the iterative solution of Gorkov's equations are compared.
The dependence of the results on the size of the single-particle model space, i.e. on the basis used to represent the matrix elements of one and two-body operators at play, is investigated in Sec. \ref{sec:mod_sp}, followed by final remarks in Sec. \ref{sec:conclusions}.

\section{Gorkov-Green's function theory}
\label{sec:GGFtheory}

\subsection{Gorkov's equation}
\label{subsec:gkv_eq}

Given the intrinsic Hamiltonian 
\begin{equation}
\label{eq:H}
H_{\text{int}}\equiv T+V-T_{CM} \: \: ,
\end{equation}
Gorkov-SCGF theory targets the ground state $|  \Psi_0 \rangle$ of the grand-canonical-like potential $\Omega \equiv H_{\text{int}} - \mu_p \, \hat{Z} -\mu_n \, \hat{N}$, having the targeted proton $Z = \langle  \Psi_0 |  \hat{Z} |  \Psi_0 \rangle$ and neutron $N = \langle  \Psi_0 |  \hat{N} |  \Psi_0 \rangle$ numbers on average. Here, $\mu_p$  ($\mu_n$) denotes the proton (neutron) chemical potential and $\hat{Z}$ ($\hat{N}$) the proton- (neutron-)number operator.

The complete dynamics is embodied in a set of four Green's functions known as Gorkov's propagators~\cite{Gorkov:1958}%
\footnote{Two-dimensional matrices in Gorkov space are denoted in boldface throughout the paper. Non-boldface quantities are used for vectors and matrices defined on the one-body Hilbert space ${\cal H}_{1}$. Matrix elements of the latter are denoted by latin letter subscripts $\{a, b, \ldots \}$, which label single-particle basis states of  ${\cal H}_{1}$.}
\begin{equation}
\label{eq:Gprop}
\mathbf{G}(\omega) = 
  \left( \begin{array}{cc}
          G^{11}(\omega) & G^{12}(\omega) \\
          G^{21}(\omega) & G^{22}(\omega) 
          \end{array}  \right) \, ,
\end{equation}
whose matrix elements read in the Lehmann representation as
\begin{subequations}
\label{eq:leh}
\begin{eqnarray}
\label{eq:leh11}
G^{11}_{ab} (\omega) &=&  \sum_{k} \left\{
\frac{\mU_{a}^{k} \,\mU_{b}^{k*}}
{\omega-\omega_{k} + \ii \eta}
+ \frac{\bar{\mV}_{a}^{k*} \, {\bar{\mV}_{b}^{k}}}{\omega+\omega_{k} - \ii \eta} \right\} \: ,\\
\label{eq:leh12}
G^{12}_{ab} (\omega) &=&   \sum_{k}
\left\{
\frac{\mU_{a}^{k} \,\mV_{b}^{k*}}
{\omega-\omega_{k} + \ii \eta} + \frac{\bar{\mV}_{a}^{k*} \, {\bar{\mU}_{b}^{k}}}{\omega+\omega_{k} - \ii \eta}
\right\}  \, ,\\
\label{eq:leh21}
G^{21}_{ab} (\omega) &=&  \sum_{k}
\left\{
\frac{\mV_{a}^{k} \,\mU_{b}^{k*}}
{\omega-\omega_{k}  + \ii \eta}
+ \frac{\bar{\mU}_{a}^{k*} \, {\bar{\mV}_{b}^{k}}}{\omega+\omega_{k} - \ii \eta}
\right\} \, ,\\
\label{eq:leh22}
G^{22}_{ab} (\omega) &=&   \sum_{k} \left\{
\frac{\mV_{a}^{k} \,\mV_{b}^{k*}}
{\omega-\omega_{k} + \ii \eta}
+  \frac{\bar{\mU}_{a}^{k*} \, {\bar{\mU}_{b}^{k}}}{\omega+\omega_{k} - \ii \eta} \right\} \: .
\end{eqnarray}
\end{subequations}
The poles of the propagators are given by $\omega_{k} \equiv \Omega_k - \Omega_0$, where the index $k$ refers to normalized eigenstates of $\Omega$ over Fock space
\begin{equation}
\label{eq:kapp}
\Omega \, | \Psi_{k} \rangle = \Omega_{k} \, | \Psi_{k} \rangle \: .
\end{equation}
The residue of  $\mathbf{G}(\omega)$ associated with pole $\omega_{k}$ relates to the probability amplitudes 
$\mU^k$ ($\mV^k$) to reach state $| \Psi_{k} \rangle$ by adding (removing) a nucleon to (from) $| \Psi_{0} \rangle$
on a single-particle state%
\footnote{The component of vector $\mU^k$ associated with a single-particle state $a$ is denoted by $\mU_{a}^{k}$. Correspondingly, the component associated with the time-reversed state $\bar{a}$ is denoted by $\bar{\mU}_{a}^{k}$~\cite{Soma:2011GkvI}.}. 

Dressed one-body propagators (Eqs.~(\ref{eq:leh})) are solutions of Gorkov's equation of motion
\begin{equation}
\label{eq:gorkov}
\left.
\left(
\begin{tabular}{c}
\hspace{-0.2cm} $T  + \Sigma^{11}(\omega)- \mu_{q_k}   \qquad \quad  \Sigma^{12}(\omega)$ \\
$\Sigma^{21}(\omega) \qquad \quad  \hspace{-0.4cm}  -T + \Sigma^{22}(\omega) + \mu_{q_k} $
\end{tabular}
\right)
\right|_{\omega_k}
\hspace{-0.2cm}
\left(
  \begin{array}{c}
\hspace{-0.1cm} \mU^k   \\
\hspace{-0.1cm} \mV^k
  \end{array} \hspace{-0.1cm} \right)
  \hspace{-0.1cm}
= \omega_{k}
\left(
  \begin{array}{c}
\hspace{-0.1cm} \mU^k   \\
\hspace{-0.1cm} \mV^k 
  \end{array} \hspace{-0.1cm} \right)  ,
\end{equation}
whose output is the set of vectors $(\mU^k, \mV^k)$ and energies $\omega_{k}$. The chemical potential $\mu_{q_k}$ is equal to $\mu_p$ or $\mu_n$  depending on the charge quantum number $q_k$ carried by the pole~$k$. Equation~(\ref{eq:gorkov}) reads as a one-body eigenvalue problem in which the normal [$\Sigma^{11}(\omega)$ and $\Sigma^{22}(\omega)$] and anomalous [$\Sigma^{12}(\omega)$ and $\Sigma^{21}(\omega)$] irreducible self-energies act as {\it energy-dependent} potentials.  Notice that $\Sigma^{11}(\omega)$ is also identified with the microscopic nucleon-nucleus optical potential~\cite{Capuzzi96,Waldecker:2011by}, allowing for the computation of scattering states~\cite{Barbieri:2005NAscatt}.

For a detailed discussion on the computation of observables, we refer the reader to Ref.~\cite{Soma:2011GkvI}. Let us limit ourselves here to defining quantities that will effectively appear in the various figures below. The total binding energy of the A-body system is computed via the Koltun sum rule~\cite{Koltun:1972}
\begin{eqnarray}
\label{eq:koltun_gorkov}
E^{\text{A}}_0 &=&
\frac{1}{4 \pi i} \int_{C \uparrow} d \omega \, \text{Tr}_{{\cal H}_{1}}\!\left[ G^{11} (\omega) \left[ T
+ \omega \right] \right] \, ,
\end{eqnarray}
which is exact for two-body Hamiltonians. Separation energies between the A-body ground state and eigenstates of $\text{A}\pm1$ systems are related to the poles $\omega_k$ through
\begin{eqnarray}
E_k^{\pm} \equiv  \mu_{q_k} \pm \omega_k 
&=&   \pm \left[\langle \Psi_k | H_{\text{int}} |  \Psi_k \rangle - \langle  \Psi_0 | H_{\text{int}} |  \Psi_0 \rangle\right] 
\nonumber    \\
&& \mp \mu_{q_k} \left[\langle  \Psi_k | \hat{Z} + \hat{N} |  \Psi_k \rangle - (\text{A} \pm 1)\right] \, ,  ~~~ 
\label{eq:epmk1}
\end{eqnarray}
where the second bracket takes care of the error associated with the difference between the average number of particles in $|  \Psi_k \rangle$ and the targeted particle number $\text{A}\pm1$. The spectral function associated with the {\it direct} addition or removal of a nucleon is then obtained according to
\begin{eqnarray}
\label{eq:sfunc2_11}
S (E) &=& \sum_{k} \mU^k \mU^{k \, \dagger} \, \delta(E-E^+_{k}) + \mV^{k \, \ast} \mV^{k \, T}  \, \delta(E-E^-_{k})  \, , \nonumber
\end{eqnarray}
from which the spectral strength distribution  (SSD) is extracted through ${\cal S}p (E) \equiv \text{Tr}_{{\cal H}_1} \left[S (E)\right] $, i.e.
\begin{eqnarray}
\label{eq:SSDan}
{\cal S}p (E) &=& \sum_k  SF_{k}^{+} \, \delta(E-E^+_k) + SF_{k}^{-} \, \delta(E-E^-_k) \: ,
\end{eqnarray}
where
\begin{subequations}
\begin{eqnarray}
SF_{k}^{+} \hspace{-0.1cm} &\equiv& \text{Tr}_{{\cal H}_{1}}\!\left[\mU^k \mU^{k \, \dagger}\right]  \, , \\
SF_{k}^{-} \hspace{-0.1cm} &\equiv& \text{Tr}_{{\cal H}_{1}}\!\left[\mV^{k \, \ast} \mV^{k \, T}\right] \, ,
\label{eq:SF}
\end{eqnarray}
\end{subequations}
define spectroscopic factors. The SSD provides the probability to leave the system  with the relative energy $E$ when adding/removing a nucleon to/from $| \Psi_0 \rangle$. Last but not least, effective single-particle energies (ESPEs) introduced by Baranger as centroids $e^{\text{cent}}_a$ of one-nucleon addition and removal spectra $E_k^{\pm}$ can be naturally computed in the present context as the eigenvalues of the first moment of the spectral function~\cite{Soma:2011GkvI,Duguet2011}.


\subsection{Self-energy expansion}
\label{subsec:se-expansion}

The solution of eigenvalue problem~(\ref{eq:gorkov}) yields the complete set of $\{\mU^k, \mV^k, \omega_{k}\}$ from which one can reconstruct Gorkov's propagators. This requires the knowledge of the self-energy, which can always be written as the sum of a static (i.e. energy independent) contribution and a dynamical term, i.e.
\begin{eqnarray}
\mathbf{\Sigma}(\omega) \equiv \mathbf{\Sigma}^{(\infty)} + \mathbf{\Sigma}^{(dyn)}(\omega) \, .
\label{eq:Sigma}
\end{eqnarray}
The four static self-energies read~\cite{Soma:2011GkvI}
\begin{subequations}
\label{eq:lambda_h}
\begin{eqnarray}
\Sigma^{11 \, (\infty)}_{ab} &=& +\sum_{cd}\bar{v}_{acbd} \, \rho_{dc}
\equiv +\Lambda_{ab}
= + \Lambda_{ab}^{\dagger}
 \\  \displaystyle
\Sigma^{22 \, (\infty)}_{ab} &=&  -\sum_{cd}\bar{v}_{\bar{b}d\bar{a}c} \, \rho_{cd}^*
= - \Lambda_{\bar{a}\bar{b}}^*  \, ,
 \\  \displaystyle
\label{eq:h_tilde}
\Sigma^{12 \, (\infty)}_{ab} &=&   \frac{1}{2}  \sum_{cd} \bar{v}_{a\bar{b}c\bar{d}} \, \tilde{\rho}_{cd}
\equiv + \tilde{h}_{ab} \: ,
 \\  \displaystyle
\label{eq:h_tilde_dagger}
\Sigma^{21 \, (\infty)}_{ab} &=& \frac{1}{2}  \sum_{cd} \bar{v}_{b\bar{a}c\bar{d}}^* \,  \tilde{\rho}_{cd}^* = + \tilde{h}_{ab}^{\dagger}
\: ,
\end{eqnarray}
\end{subequations}
where $\bar{v}_{acbd}$ denote antisymmetrized matrix elements of the two-body interaction entering Eq.~\eqref{eq:H}, whereas
$\rho_{ab}$ and $\tilde{\rho}_{ab}$ are respectively the normal and anomalous density matrices 
\begin{subequations}
\label{eq:allbdm}
\begin{eqnarray}
\label{eq:nobdm}
\rho_{ab} &\equiv& \langle \Psi_0 | a_b^{\dagger} a_a | \Psi_0 \rangle  
= \sum_{k} \bar{\mV}_{b}^k \, \bar{\mV}_{a}^{k*} \: , \\
\label{eq:aobdm}
\tilde{\rho}_{ab} &\equiv& \langle \Psi_0 | \bar{a}_b a_a | \Psi_0 \rangle
= \sum_{k} {\bar{\mU}_{b}^k} \, \bar{\mV}_{a}^{k*}   \: .
\end{eqnarray}
\end{subequations}
Equations~(\ref{eq:lambda_h}) are formally of first order in $V$ and resemble plain Hartree-Fock and Bogoliubov one-body fields. However, they are expressed in terms of {\it fully correlated} ground-state density matrices $\rho$ and $\tilde{\rho}$. Thus, they implicitly sum all {\it static} higher-order diagrams in perturbation theory. In the presence of three- or higher many-body interactions, they acquire further contributions due to additional interaction reducible diagrams~\cite{Carbone:2013eqa}.

If only first-order contributions to the self-energy are actually retained, Eqs.~(\ref{eq:gorkov}), (\ref{eq:lambda_h}) and (\ref{eq:allbdm}) do reduce to an \textit{ab initio} Hartree-Fock-Bogoliubov (HFB) problem. At higher orders, the self-energy acquires energy dependent contributions and the solution of Eq.~(\ref{eq:gorkov}) complicates. 
The dynamical part of the self-energy can be expressed through its Lehmann representation as follows
\begin{subequations}
\label{eq:leh_self}
\begin{eqnarray}
\label{eq:leh_self11}
\Sigma_{ab}^{11 \, (dyn)} (\omega) &=&
\sum_{\kappa}  \left\{
\frac{{\mC^{\kappa}_{a}}
\, ({\mC^{\kappa}_{b}})^{*}}{\omega-E_{\kappa} + \ii \eta}
+ \frac{({\bar{\mD}^{\kappa}_{a}})^{*}
\, \bar{\mD}^{\kappa}_{b}}{\omega+E_{\kappa} - \ii \eta}
\right\}  , \quad\quad
\\ \displaystyle
\label{eq:leh_self12}
\Sigma_{ab}^{12 \, (dyn)} (\omega) &=&
\sum_{\kappa}  \left\{
\frac{{\mC^{\kappa}_{a}}
\, ( {\mD}^{\kappa}_{b}  )^* }{\omega-E_{\kappa} + \ii \eta}
+ \frac{({\bar{\mD}^{\kappa}_{a}})^{*}
\, {\bar{\mC}^{\kappa}_{b}}}{\omega+E_{\kappa} - \ii \eta}
\right\} , \quad
\\ \displaystyle
\label{eq:leh_self21}
\Sigma_{ab}^{21 \, (dyn)} (\omega) &=&
\sum_{\kappa}  \left\{
\frac{{\mD^{\kappa}_{a}}
\, ({\mC^{\kappa}_{b}})^{*}}{\omega-E_{\kappa} + \ii \eta}
+ \frac{({\bar{\mC}^{\kappa}_{a}})^*
\, \bar{\mD}^{\kappa}_{b}}{\omega+E_{\kappa} - \ii \eta}
\right\} , \quad
 \\ \displaystyle
\label{eq:leh_self22}
\Sigma_{ab}^{22 \, (dyn)} (\omega) &=&
\sum_{\kappa}  \left\{
\frac{{\mD^{\kappa}_{a}}
\, (\mD^{\kappa}_{b})^*}{\omega-E_{\kappa} + \ii \eta}
+ \frac{({\bar{\mC}^{\kappa}_{a}})^*
\, {\bar{\mC}^{\kappa}_{b}}}{\omega+E_{\kappa} - \ii \eta}
\right\} , \quad
\end{eqnarray}
\end{subequations}
where $\mC$ and $\mD$ account for the coupling of one quasi-particle excitations to configurations involving $2n+1$ quasi-particles, with $n\geq1$, while  $E_{\kappa}$ labels the energy of such configurations.  The structure of Eqs.~(\ref{eq:leh_self}) does not change if additional many-body interactions enter the Hamiltonian. Up to this point no approximation has been made, i.e. if the exact self-energy is employed in Eqs~(\ref{eq:lambda_h}) and~(\ref{eq:leh_self}) then Gorkov's Eq.~(\ref{eq:gorkov}) is equivalent to solving the exact \hbox{$A$-body} Schr\"odinger equation.

In actual calculations, a truncation in the expansion of $\mathbf{\Sigma}(\omega)$ has to be adopted to approximate the coupling amplitudes ($\mC^\kappa$, $\mD^\kappa$) and their poles $E_\kappa$. In the present work first- and second-order self-energy contributions are considered~\cite{Soma:2011GkvI}. Summing the eight second-order skeleton diagrams expressed in terms of correlated propagators, one obtains an approximation for $\mathbf{\Sigma}^{(dyn)}(\omega)$ with the same form of Eqs.~(\ref{eq:leh_self}) where the label $\kappa$ runs over all possible three-quasiparticles (3QP) excitations $\kappa = \{ k_1, k_2, k_3 \}$.
The corresponding poles are
\begin{equation}
\label{eq:E2nd}
E_{\kappa} = E_{k_1 k_2 k_3} \equiv \omega_{k_1} + \omega_{k_2} + \omega_{k_3} 
\end{equation} \\
while the coupling amplitude read 
\begin{subequations}
\label{eq:CD2nd}
\begin{eqnarray}
\mC^{k_1k_2k_3}_{a} &\equiv& \frac{1}{\sqrt{6}} \left [ {\mM^{k_1k_2k_3}_{a}} + {\mM^{k_2k_3k_1}_{a}} + {\mM^{k_3k_1k_2}_{a}} \right ]  ,\quad
\\
\mD^{k_1k_2k_3}_{a} &\equiv& \frac{1}{\sqrt{6}} \left [ {\mN^{k_1k_2k_3}_{a}} + {\mN^{k_2k_3k_1}_{a}} + {\mN^{k_3k_1k_2}_{a}} \right ] , \quad
\end{eqnarray}
\end{subequations}
where
\begin{subequations}
\label{eq:mpr}
\begin{eqnarray}
\mM^{k_1k_2k_3}_{a} &\equiv& \sum_{ijk} \bar{v}_{akij} \,\mU_{i}^{k_1} \mU_{j}^{k_2} \bar{\mV}_{k}^{k_3} \, , \\
\mN^{k_1k_2k_3}_{a} &\equiv& \sum_{ijk} \bar{v}_{akij} \, \mV_{i}^{k_1} \mV_{j}^{k_2} \bar{\mU}_{k}^{k_3} \, .
\end{eqnarray}
\end{subequations}

\begin{widetext}
\subsection{Energy-independent form of Gorkov's equation}
\label{subsec:energy}

Using Eq.~\eqref{eq:leh_self}, an alternative formulation of Gorkov's equation can be derived. Introducing the two additional amplitudes $\mW$ and $\mZ$ that describe the admixtures of 3QP configurations according to
\begin{subequations}
\label{eq:wz}
\begin{eqnarray}
(\omega_k-E_{k_1 k_2 k_3}) \, \mW^{k_1k_2k_3}_{k} &\equiv&
\sum_a \left[ ({\mC^{k_1k_2k_3}_{a}})^{*} \, \mU^k_a + ({\mD^{k_1k_2k_3}_{a}})^{*} \, \mV^k_a \right] \: ,
\\
(\omega_k+E_{k_1 k_2 k_3}) \, \mZ^{k_1k_2k_3}_{k} &\equiv&
\sum_a \left[  \bar{\mD}^{k_1k_2k_3}_{a} \, \mU^k_a + {\bar{\mC}^{k_1k_2k_3}_{a}} \, \mV^k_a \right] \: ,
\end{eqnarray}
\end{subequations}
Eq.~\eqref{eq:gorkov} can be rewritten as
\begin{subequations}
\label{eq:gorkov_premat}
\begin{eqnarray}
\omega_k \, \mU^k_a &=& \sum_b \left [ (T_{ab} - \mu \, \delta_{ab} + \Lambda_{ab}) \, \mU^k_b + \tilde{h}_{ab} \, \mV^k_b \right]
+ \sum_{k_1k_2k_3} \left [ {\mC^{k_1k_2k_3}_{a}} \, \mW^{k_1k_2k_3}_{k}
+ ({\bar{\mD}^{k_1k_2k_3}_{a}})^{*} \, \mZ^{k_1k_2k_3}_{k} \right]  \: ,
\\
\omega_k \, \mV^k_a &=& \sum_b \left [\tilde{h}_{ab}^{\dagger} \, \mU^k_b  -(T_{ab} - \mu \, \delta_{ab} + \Lambda_{\bar{a}\bar{b}}^*) \, \mV^k_b\right]
+ \sum_{k_1k_2k_3} \left [{\mD^{k_1k_2k_3}_{a}} \, \mW^{k_1k_2k_3}_{k}
+ ({\bar{\mC}^{k_1k_2k_3}_{a}})^{*} \, \mZ^{k_1k_2k_3}_{k} \right]  \: .
\end{eqnarray}
\end{subequations}
The four relations above provide a set of coupled equations for unknowns $\mU$, $\mV$, $\mW$ and $\mZ$ that can be recast in a matrix form
\begin{equation}
\label{eq:xi}
\Xi
\left(
\begin{array}{c}
\mU  \\ \mV \\ \mW \\ \mZ
\end{array}
\right)_k \equiv
\left(
\begin{array}{cccc}
h &  \tilde{h} &\quad  \mC \ &\quad  \bar{\mD}^{*}  \\
\tilde{h}^{\dagger} & -\bar{h}^* &\quad  \mD &\quad  \bar{\mC}^* \\
\mC^{\dagger} & \mD^{\dagger}  &\quad  E  &\quad  0 \\
\bar{\mD}^{T} & \bar{\mC}^{T}  &\quad  0 &\; \; -E
\end{array}
\right)
\left(
\begin{array}{c}
\mU  \\ \mV \\ \mW \\ \mZ
\end{array}
\right)_k
= \omega_k
\left(
\begin{array}{c}
\mU  \\ \mV \\ \mW \\ \mZ
\end{array}
\right)_k
\: ,
\end{equation}
\end{widetext}
where $h \equiv T-\mu+\Lambda$ and $E \equiv \text{diag} \{ E_\kappa \}$. The derivation of the energy-independent matrix $\Xi$ can be generalized to higher-order truncations of the self-energy as long as the latter can be expressed through the Lehmann representation~\eqref{eq:leh_self}.

\section{Numerical algorithm}
\label{sec:numerical}

Identical solutions are associated with Gorkov's equation in the form \eqref{eq:gorkov} or \eqref{eq:xi}. 
Numerically, however, the treatment of an energy-dependent eigenvalue equation is not particularly desirable. 
Attempts solve Eq.~\eqref{eq:gorkov} directly have revealed problematic due to the presence of the energy denominators in  ${\bf \Sigma}(\omega)$ that imply drastic variations of the self-energy near its poles~\cite{Bergli2011}.  Even with very fine meshes in energy, this issue  severely limits the resolution of the calculation~\cite{Brand1988}. Alternatively, each pole can be searched for individually~\cite{YuanPhD1994,Barbieri2002,Barbieri:2007Atoms} but this involves a lengthy numerical procedure  that does not guarantee the access to all solutions of  Eq. \eqref{eq:gorkov}, i.e. a sizeable fraction of the spectral strength may be neglected.
Working with Eq. \eqref{eq:xi}, on the other hand, avoids divergences and automatically guarantees the extraction of all the poles at once. The price to pay is a severe growth in the dimension of Gorkov's matrix, with consequent limitations on its diagonalization and a stringent requirement in memory storage. Nevertheless, this eventually results in a gain of more than one order of magnitude in computational time with respect to solving Eq.~\eqref{eq:gorkov} directly. As discussed at length in the following, the large dimension of $\Xi$ does  not preclude convergence in model spaces that are large enough for modern \textit{ab initio} nuclear structure calculations.

\subsection{Self-consistency and dimensionality}
\label{sec:sc-dim}

\begin{figure}[h]
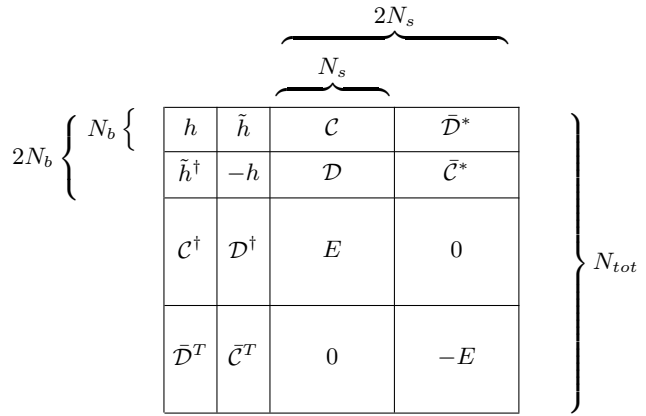

$$
\begin{tabular}{m{1.25cm}m{0.55cm}|m{0.55cm}|m{0.55cm}|m{1.5cm}|m{1.5cm}|c} 
\multicolumn{4}{c}{}  & \multicolumn{2}{c}{$\overbrace{\hspace{3cm}}^{\mbox{\small{$2N_s$}}}$} &  \\ 
\multicolumn{4}{c}{}   & \multicolumn{1}{c}{$\overbrace{\hspace{1.5cm}}^{\mbox{\small{$N_s$}}}$} &  \multicolumn{2}{c}{} \\ 
\cline{3-6} \multirow{2}{*}{$2 N_b \left\{ \mbox{\fontsize{26}{40}\selectfont \phantom{I}} \right.$} &
\hspace{-0.5cm} $N_b \left\{ \mbox{\fontsize{16}{40}\selectfont \phantom{I}} \right.$ & \centering$h$ & \centering$\tilde{h}$ & \centering$ \mC$ & \centering$  \bar{\mD}^{*}$ & 
\multirow{4}[4]{*}{\hspace{-0.6cm} $\left. \mbox{\fontsize{88}{70}\selectfont \phantom{I}} \right\} N_{tot}$} \\
\cline{3-6}  & \mbox{\fontsize{21}{40}\selectfont \phantom{I}} & \centering$ \tilde{h}^{\dagger}$ & \centering$-h$  & \centering $\mD$ & \centering $\bar{\mC}^*$ &  \\
\cline{3-6} \multicolumn{2}{m{0.75cm}|}{$\mbox{\fontsize{55}{40}\selectfont \phantom{I}}$} 
& \centering $\mC^{\dagger}$ & \centering $\mD^{\dagger}$ & \centering $E$ & \centering 0&  \\
\cline{3-6} \multicolumn{2}{m{0.75cm}|}{$\mbox{\fontsize{55}{40}\selectfont \phantom{I}}$} 
& \centering $\bar{\mD}^{T}$ & \centering $\bar{\mC}^{T}$ & \centering 0 & \centering $-E$ &  \\
\cline{3-6}
\end{tabular}
$$
\caption{Dimension scheme for the Gorkov matrix $\Xi$.}
\label{fig:xi}
\end{figure}

Gorkov's matrix depends on eigenvalues $\omega_k$ and amplitudes ($\mU^k$,$\mV^k$), which implies that the solution must be searched for iteratively.  To see how the energy-independent form, Eq. \eqref{eq:xi}, involves a drastic increase of the dimensionality of the problem at each iteration, let us partition the matrix $\Xi$ as follows
\begin{equation}
\label{eq:xi_12}
\Xi
=
\left(
\begin{array}{cc|cc}
h &  \tilde{h} &\quad  \mC \ &\quad  \bar{\mD}^{*}  \\
\tilde{h}^{\dagger} & -\bar{h}^* &\quad  \mD &\quad  \bar{\mC}^* \\ \hline
\mC^{\dagger} & \mD^{\dagger}  &\quad  E  &\quad  0 \\
\bar{\mD}^{T} & \bar{\mC}^{T}  &\quad  0 &\; \; -E
\end{array}
\right)
\equiv
\left(
\begin{array}{c|c}
 \Xi^{(1)} & \Xi^{(2)} \\ \hline
 \Xi^{(2) \, \dagger}  &  \mathcal{E}
\end{array}
\right)
\: .
\end{equation}
The number of states in the single-particle basis, $N_b$,  defines the dimension of the first-order block $\Xi^{(1)}$ (see Fig.~\ref{fig:xi}). 
Each of the four sub-blocks in $\Xi^{(1)}$ is $N_b \times N_b$, for a total of $2 N_b \times 2 N_b$ matrix elements. The matrix $E$ is diagonal for second-order self-energies and its elements are all possible combinations of three pole energies $\{ \omega_{k_1}, \omega_{k_2}, \omega_{k_3}\}$. A product state solution of the HFB problem is typically chosen as the reference state so that
$N_b$ positive quasi-particle energies are involved at the first iteration. In this situation, the number of poles in Eqs.~(\ref{eq:leh_self}) is
\begin{equation}
N_s \approx \left( 
\begin{array}{c}
N_b \\
3
\end{array}
\right) 
\approx \frac{N_b^3}{6}
\: .
\end{equation}
Since $N_b \ll (N_b)^3$ it follows that dim$(\Xi) = N_{tot} \approx N_b^3/3$. In a general, e.g. m-scheme, implementation $N_b$ of order of a few hundreds is typically needed to achieve convergence. Thus, the diagonalization of Gorkov's matrix  for large model spaces may be infeasible with current computational resources, even for the first iteration.

Diagonalizing $\Xi$ the first time, about $(N_b)^3/6$ new poles (i.e. one quasi-particle states) are generated, which represent the new fragments carrying each a fraction of the spectral strength distribution. In the second iteration, the number of possible three quasi-particle energies $E_{k_1 k_2 k_3}$ has increased accordingly, resulting in $N_s \approx N_b^9/216/6$, which leads to dim$(\Xi) \approx N_b^9/1000 \times N_b^9/1000$. In the $n$-th iteration the matrix $\Xi$ will have expanded to dimensions of order $N_b^{3^n} \times N_b^{3^n}$. This growth clearly prevents the exact treatment of all poles in an actual (self-consistent) calculation and one has to keep dim($\Xi$) below a threshold that makes the scheme computationally tractable.

\subsection{Krylov projection}
\label{subsec:krylov}

We follow Ref.~\cite{Schirmer1989} and project the energy denominators of  ${\bf \Sigma}^{(dyn)}(\omega)$ to a smaller Krylov subspace. Doing so, the dimensional growth of Gorkov's matrix is contained and a sustainable computational procedure can be developed. 

We consider a set of pivot vectors  $p^i$ with elements
\begin{equation}
\label{eq:pivots}
 p^i_\kappa = \sum_a \, \mC_a^\kappa  U^i_a \; + \; \sum_a \, \mD_a^\kappa V^i_a 
\: ,
\end{equation}
where ($U^i$, $V^i$) are linearly independent vectors in the space of HFB quasi-particle states, i.e of the $2N_b$ eigensolutions of $\Xi^{(1)}$.  In general, one needs as many pivots as there are single-particle basis states in the model space to properly converge  all off-diagonal elements of Eqs. \eqref{eq:leh_self}~\cite{Schirmer1989}. Up to $N_p=2 N_b$ starting pivots are thus used to generate a Krylov subspace $\mathcal{K}$ associated with the submatrix $E$ in Eq.~(\ref{eq:xi}).
Our particular implementation uses a Lanczos-type algorithm that uses one pivot at a time and iterates it $N_\ell$ times, independently of the others.  Each time Lanczos iterations are started with a new pivot, $p^i$, it is first orthogonalised with respect to the basis vectors already generated.  This is equivalent to a block Lanczos reduction based on a slightly modified set of pivots $\{p^i{}'\}$. Eventually, the dimension of the Krylov space is the number of \textit{total} Lanczos iterations, $N_L=$ dim$(\mathcal{K})= N_\ell \times N_{p}$.
Full details of the algorithm are given in Appendix~\ref{app:lanczos}.

The block $\mathcal{E}$ in Eq.~(\ref{eq:xi_12}) reduces to a matrix of lower dimensions
\begin{equation}
\label{eq:Krylov_E}
\mathcal{E} \longrightarrow \mathcal{E}' =
\left(
\begin{array}{cc}
   \mL^{\dagger}  \, E \, \mL   \\
   & - \mL^{\dagger}  \, E \, \mL
\end{array}
\right)
\: ,
\end{equation}
where $\mL$ is the collection of vectors generated by the Lanczos procedure. 
The two off-diagonal blocks $\Xi^{(2)}$ and $\Xi^{(2) \, \dagger}$ are transformed accordingly:
\begin{subequations}
\label{eq:Krylov_Xi}
\begin{eqnarray}
\Xi^{(2)} &\longrightarrow& \Xi'^{(2)} = \Xi^{(2)} \, \left( \begin{array}{c} \mL \\ \mL \end{array} \right) \:\: , \\
\Xi^{(2)\, \dagger} &\longrightarrow& \Xi'^{(2)\, \dagger} =   \left( \begin{array}{cc} \mL^\dagger &  \mL^\dagger \end{array} \right)  \, \Xi^{(2)\, \dagger}  \:\: .
\end{eqnarray}
\end{subequations}
These projected blocks are inserted in the original Gorkov matrix
\begin{equation}
\label{eq:xi_12prime}
\Xi \longrightarrow \Xi'
=
\left(
\begin{array}{c|c}
 \Xi^{(1)}                  &     \Xi'^{(2)} \\ \hline
 \Xi'^{(2)\, \dagger}  &    \mathcal{E}' 
\end{array}
\right)
\: ,
\end{equation}
whose dimension is now dim$(\Xi') = N'_{tot} \times N'_{tot}  = (2 N_b + 2 N_L) \times (2 N_b + 2 N_L)$.  Gorkov-Krylov's matrix $\Xi'$ is finally (fully) diagonalized with standard diagonalization routines. For a sufficiently large number of iterations \hbox{dim$(\mathcal{K})$ $\rightarrow$ dim$(E)$} and the exact result is recovered. In terms of Lehmann representation, Eq.~\eqref{eq:leh_self},   the Krylov projected quantities results in approximating the exact self-energy as 
\begin{equation}
\Sigma^{(2)} = \mC \frac{1}{\mathbb{1} \omega - E} \mC^{\dagger} \:\: \longrightarrow \:\:
\Sigma'^{(2)} = \mC \, \mL \frac{1}{\mathbb{1} \omega - \mL^{\dagger} \, E \, \mL} \mL^{\dagger} \, \mC^{\dagger} \: ,
\label{eq:Krylov_prop}
\end{equation}
where only the first term in Eq.~\eqref{eq:leh_self11} has been considered for illustration. The other terms follow accordingly.

The technique outlined here differs in spirit from the standard use of Lanczos or Arnoldi algorithms in large-scale
shell-model diagonalizations. While these methods aim at excellent estimates of the lowest eigenvalues of a large matrix, in SCGF calculations one is also interested in reproducing most of the key features of the spectral distribution. 
The Krylov projection of a matrix ensures a fast convergence at the extremes of its eigenvalue spectrum. Thus, it is important that the Lanczos algorithm is applied separately to both sub-blocks $E$  and $-E$ of Eq.~(\ref{eq:xi_12}), which are mirrored across the Fermi energy. In this way the quasiparticle spectrum near the Fermi surface is recovered accurately upon diagonalizing Eq.~(\ref{eq:xi_12prime}). The other important property of Krylov projection techniques is that the  first $2 N_\ell$ moments  of each pivot are conserved during the projection. This ensures that the overall SSD converges quickly, which is important for achieving good estimates for all observables after a relatively small number of Lanczos iterations.

\subsection{Calculation scheme}
\label{subsec:calculation_scheme}

To obtain the self-consistent solution for the four Gorkov propagators, the following steps are performed
\begin{enumerate}
\item Reference propagators are used as an initial set of $\{ \mU^k, \mV^k, \omega_k \}$ and $\{ \mu_p, \mu_n \}$. They are typically generated by solving the first-order HFB eigenvalue problem.
\item Second-order self-energies are computed through Eqs.~\eqref{eq:E2nd} to~\eqref{eq:mpr}. 
\item The Krylov projection is performed according to Eqs.~\eqref{eq:Krylov_E} and~\eqref{eq:Krylov_Xi}.
\item Energy-independent self-energies entering $\Xi^{(1)}$ are computed by means of Eqs. \eqref{eq:lambda_h}.
\item Matrix $\Xi'$ [Eq. \eqref{eq:xi_12prime}] is constructed and diagonalized.
\item Chemical potentials $\mu_p$ and $\mu_n$ are adjusted to yield on average the proton and neutron numbers of the targeted nucleus according to Eq. \eqref{eq:adjust_mu}. This involves several re-diagonalizations of matrix $\Xi'$ along with repeated adjustments of $\mu_p$ and $\mu_n$.
\item The solution (i.e. a new set of $\{ \mU^k, \mV^k, \omega_k \}$ and $\{ \mu_p, \mu_n \}$) provides updated Gorkov's propagators and is used as an input to the next iteration. The procedure re-starts from point 2 for full self-consistency (or from point 4 for the partial ``sc0'' scheme discussed below).
\end{enumerate}

The above procedure is repeated until convergence is achieved. The convergence is typically assessed by looking at the variation of the chemical potentials and/or of the total binding energy. In the present work the convergence criterion is set by variations in the total energy that are smaller than 1~keV. As discussed in the next section, such a value is smaller than the systematic error induced by the numerical algorithm.

Repeating points 2--6 above provides the fully self-consistent (``sc'') implementation of Gorkov-Green's function theory. In this case, converged results are completely independent of the reference state adopted at point 1.   A computationally cheaper alternative---referred to as ``sc0'' in the following---consists of iterating only points 4--6. In other words, self-consistency is limited to the energy independent part of the self-energy [$\Xi^{(1)}$ in Eq.~\eqref{eq:xi_12}] whereas ${\bf\Sigma}^{(dyn)}(\omega)$ is computed once and frozen afterwards. In actual calculations we employ HFB propagators to generate the second-order skeleton diagrams contributing to ${\bf\Sigma}^{(dyn)}(\omega)$. Thus, a substantial portion of self-energy insertion diagrams (beyond second order) are effectively recovered. Effectively, the partial self-consistency of the sc0 approach already retains the most important features since it implicitly generates all energy independent diagrams above first order through the dressing of propagators in Eqs.~\eqref{eq:lambda_h} and~\eqref{eq:allbdm}. As opposed to perturbation theory, the self-consistent character of Green's function methods guarantees the resummation of self-energy insertions to all orders and makes the method intrinsically non-perturbative and iterative. Since the self-energy is computed at second order in skeleton diagrams, both sc0 and sc generate all diagrams entering  ${\bf\Sigma}^{(\infty)}$ up to third order and all those entering ${\bf\Sigma}^{(dyn)}(\omega)$ up to second order.  In both cases, the self-consistency in ${\bf\Sigma}^{(\infty)}$ automatically includes all-order resummations of several diagrams beyond third order.

The two schemes will be compared in details in Sec.~\ref{sec:self}, where it will be shown that the sc0 degree of self-consistency is capable of grasping most of the correlations introduced by second-order self-energies.

\subsection{Numerical scaling}
\label{subsec:scaling}

An important issue for \textit{ab initio} approaches concerns the possibility to perform numerical calculations with increasingly large model spaces, so that it is possible to control their convergence and access heavier systems. Thus, we analyse the numerical scaling  of Gorkov-Green's function calculations. Provided that a full diagonalization of the unprojected Gorkov matrix is computationally too expensive, we directly consider the cost of calculations based on the Lanczos algorithm.

As discussed in Sec.~\ref{subsec:krylov}, the benefit of the Krylov projection regards the reduced dimensionality of Gorkov's eigenvalue problem due to the fact that $N_L \ll N_s$. Not only $\mbox{dim}(\mathcal{K}) \ll \,\mbox{dim}(E)$ but  $N_L= 2 \, N_b \times N_\ell $ is independent of the number of poles in the iterated propagator. Thus, $\mbox{dim}(\mathcal{K})$ remains small due to the Krylov projections at each iteration, allowing for self-consistent calculations even for large bases.

Before comparing the overall costs of sc0 and sc calculations, we investigate the scaling of separate steps defining the algorithm presented in Sec.~\ref{subsec:calculation_scheme}. The three main steps are (i) the calculation of Gorkov's matrix,  (ii) its Krylov reduction and (iii) its diagonalization. Only the last step is iterated in a sc0 calculation. These operations display different scaling behaviors when varying the size of model and Krylov spaces
\begin{enumerate}
\item While matrix $E$ is trivial at second order, $\Xi^{(2)}$ is made of  $2 N_b \times 2 N_s$ elements to be computed. For the first iteration and the sc0 scheme, $N_s \approx N_b^3$/6 and the elements of  $\Xi^{(2)}$ are the interaction matrix elements, Eq.~\eqref{eq:CD2nd}, expressed in the reference (HFB) basis. For successive iterations, $N_s \propto N_L^3$ while $\Xi^{(2)}$ requires projecting the interaction on Gorkov orbitals [see Eqs.~(\ref{eq:mpr})]. This requires a number of operations of order $N_b \times  N_s \times N_b^3$. Hence, calculating Gorkov's matrix scales as
%
%
\begin{enumerate}
\item $N_b^4$ for the first iteration and sc0;
\item $N_b^7 \, N_\ell^3$ for successive iterations in sc.
\end{enumerate}
%
%
\item The Lanczos algorithm iteratively generates $N_L$ basis vectors of dimension $N_s$. Within this procedure, the most time consuming part is the projection of the coupling amplitudes to obtain $\Xi'^{(2)}$ [see Eqs.~(\ref{eq:Krylov_Xi})], which is a matrix multiplication requiring $2N_b \times N_s \times N_L$ steps. Hence, the Krylov projection scales as
\begin{enumerate}
\item $N_b^5 \, N_\ell$ for the first iteration and sc0;
\item $N_b^5 \, N_\ell^4$ for successive iterations in sc.
\end{enumerate}
%
%
\item The diagonalization of Gorkov-Krylov's matrix scales as $(N'_{tot})^3 \propto N_b^3 \, N_\ell^3$, for large values of  $N_\ell$.
\end{enumerate}

\begin{figure*}[t]
\begin{center}
        \includegraphics[height=.35\linewidth]{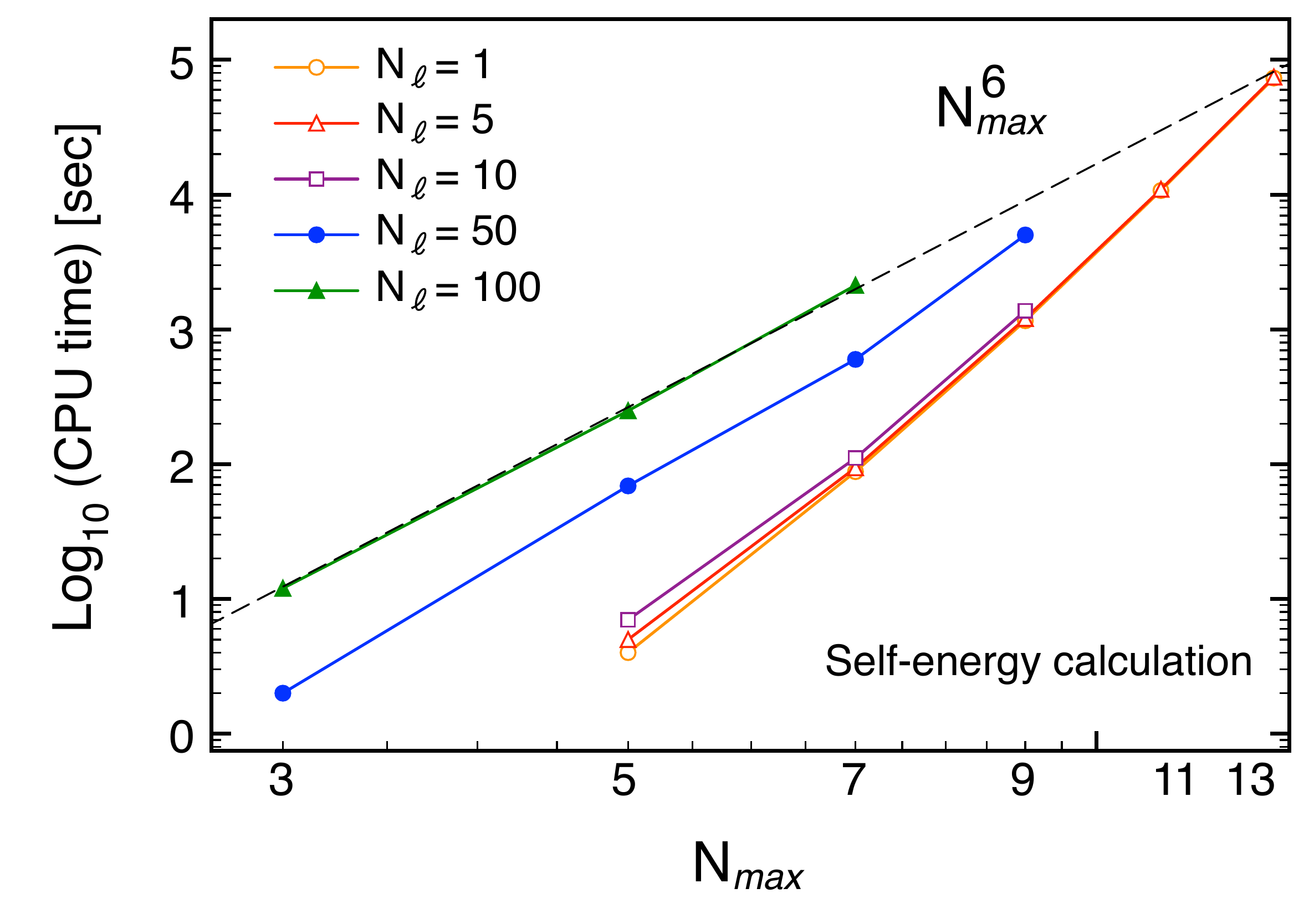}
        \includegraphics[height=.35\linewidth]{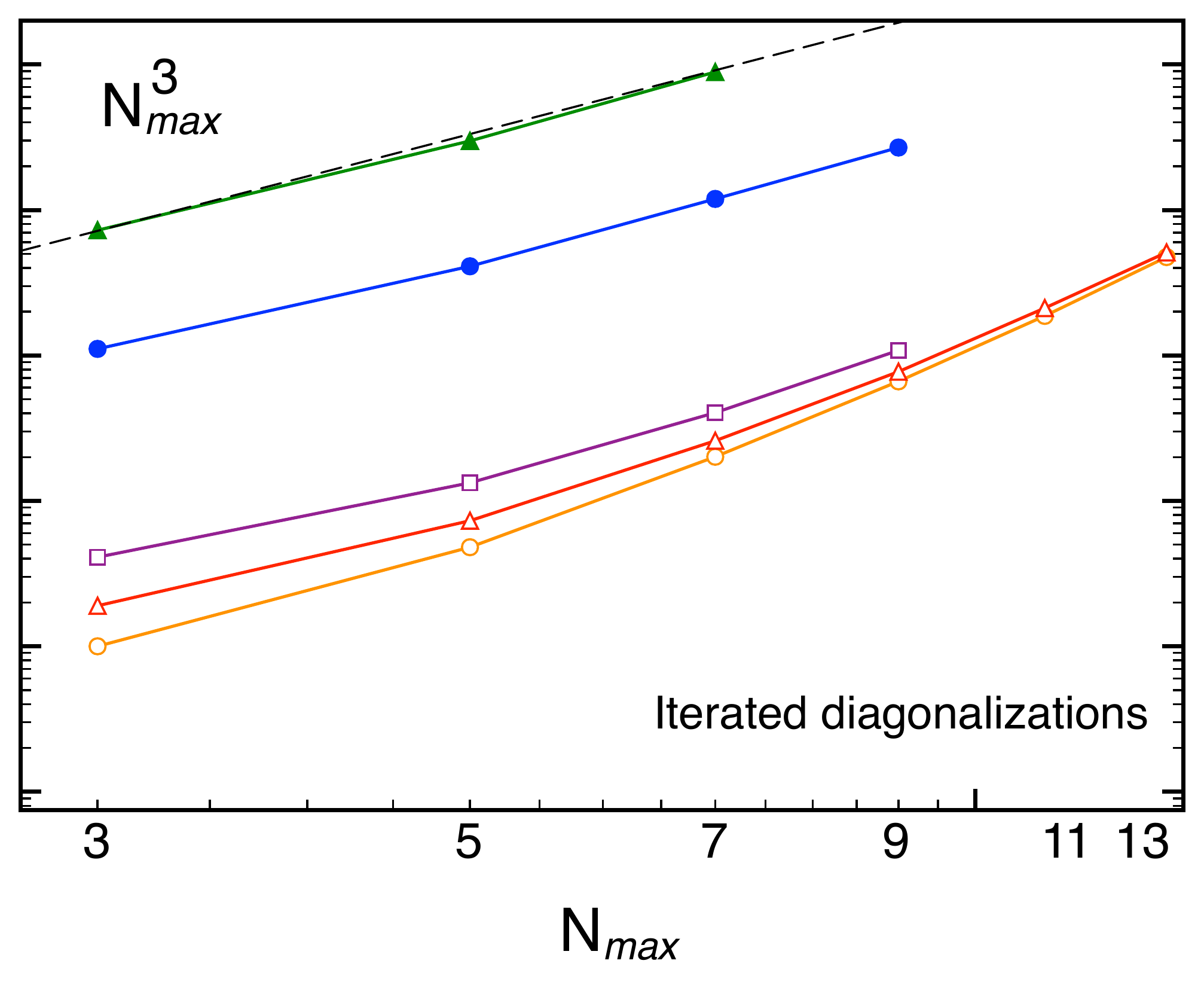}
	\caption{(Color online) CPU time spent performing specific operations during a typical sc0 calculation as a function of $\nmax$ 
	    and for different values of $N_\ell$. The contribution of the various partial waves  $\alpha$ are added.
	    Left panel: time needed to calculate the self-energies and project them
	    to Krylov's subspace (points 1 and 2 of Sec.~\ref{subsec:scaling}).
	    Right panel: time required to diagonalize Gorkov's matrix over 100 sc0 iterations (point 3 of Sec.~\ref{subsec:scaling}).
	    Dashed lines show scalings of the type  $(\nmax)^\gamma$, with $\gamma=$6 (left panel) and~3 (right panel).
	    }
\label{fig:scaling}
\end{center}
\end{figure*}

Considerations made so far are valid for a general choice of the single-particle basis $\{ a_a^{\dagger} \}$, e.g. in an m-scheme calculation, and represent a worst case scenario. Our actual implementation  
considers nuclei that are assumed to be in a $J^\Pi = 0^+$ state, for which Gorkov's equation separates
into partial waves of a given charge, angular momentum and parity, $\alpha \equiv (q, j , \pi)$~\cite{Soma:2011GkvI}.
The basis associated with a partial wave $\alpha$ has a dimension $N_b^{\alpha}$ that  corresponds 
to the number of its principal levels included in the model space. The dimension of the Krylov space, $N_L^{\alpha} = 2 \, N_b^{\alpha} \, N_\ell$, varies with $\alpha$ accordingly. This changes the above stated scaling properties in a non trivial way.
Present calculations use a spherical harmonic oscillator basis with all orbits included up to a maximum 
shell $\nmax=\max\{2n_\alpha + \ell_\alpha\}$. This basis has $N_b^\alpha \leq \nmax/2$ but the overall scaling gains an extra power in $N_b^\alpha$ because the same calculation is performed separately for each partial wave: more precisely, $\sum_\alpha (N_b^\alpha)^\gamma \propto (\nmax)^{\gamma +1}$ for large $\nmax$.
Note, however, that the relevant quantity for the m-scheme case is the total number of all possible single particle orbits, $N^{tot}_b=\sum_\alpha N_b^\alpha \propto (\nmax)^2 $. Hence, decoupling all partial waves results
in a high gain in computational time. 
In addition, for a fixed $\nmax$, the dimension of the Lanczos vectors, $N_s^\alpha$, is no longer proportional
to $(N_b^\alpha)^3$ but displays a bell shape with increasing angular momentum $j_\alpha$ that results from the combinatorics involved in coupling angular momenta. This also affects the considerations at points 1 and 2 above and results in a more gentle scaling. 

Since the actual scaling in computer time depends non trivially on the model space chosen, we tested our code 
directly in actual calculations. The results are shown in Fig.~\ref{fig:scaling} for a series of sc0 runs on a single processor. 
Left panel shows the time required to generate the Gorkov-Krylov matrix (steps 1 and 2 above) for different model-space sizes and values of $N_\ell$. For large $N_\ell$ the computation time is dominated by the Lanczos procedure and scales as $\nmax^6$, as expected.
The calculation of the second order self-energy (step 1) is significant only when using very few Lanczos iterations (when step 2 is negligible). However, it increases more rapidly with respect to $\nmax$, indicating that for large model spaces an improvement
of our algorithm for step 1 might be in order.
Right panel shows the time required for 100 diagonalizations of Gorkov-Krylov's matrix (steps 4-6
of Sec.~\ref{subsec:calculation_scheme}).
This is representative of the typical number of sc0 iterations needed in actual calculations to converge both the propagator
and the chemical potentials. The diagonalization of Eq.~(\ref{eq:xi_12prime}) becomes dominant
for large  $N_\ell$ and scales as $\nmax^3$. Both panels in Fig.~\ref{fig:scaling} reflect the actual computing time of a typical sc0 calculation and indicate that resources are evenly split between the Krylov projection and the sc0 iterations needed to reach self-consistency.

\begin{figure}[b]
\begin{center}
        \includegraphics[width=1.0\linewidth]{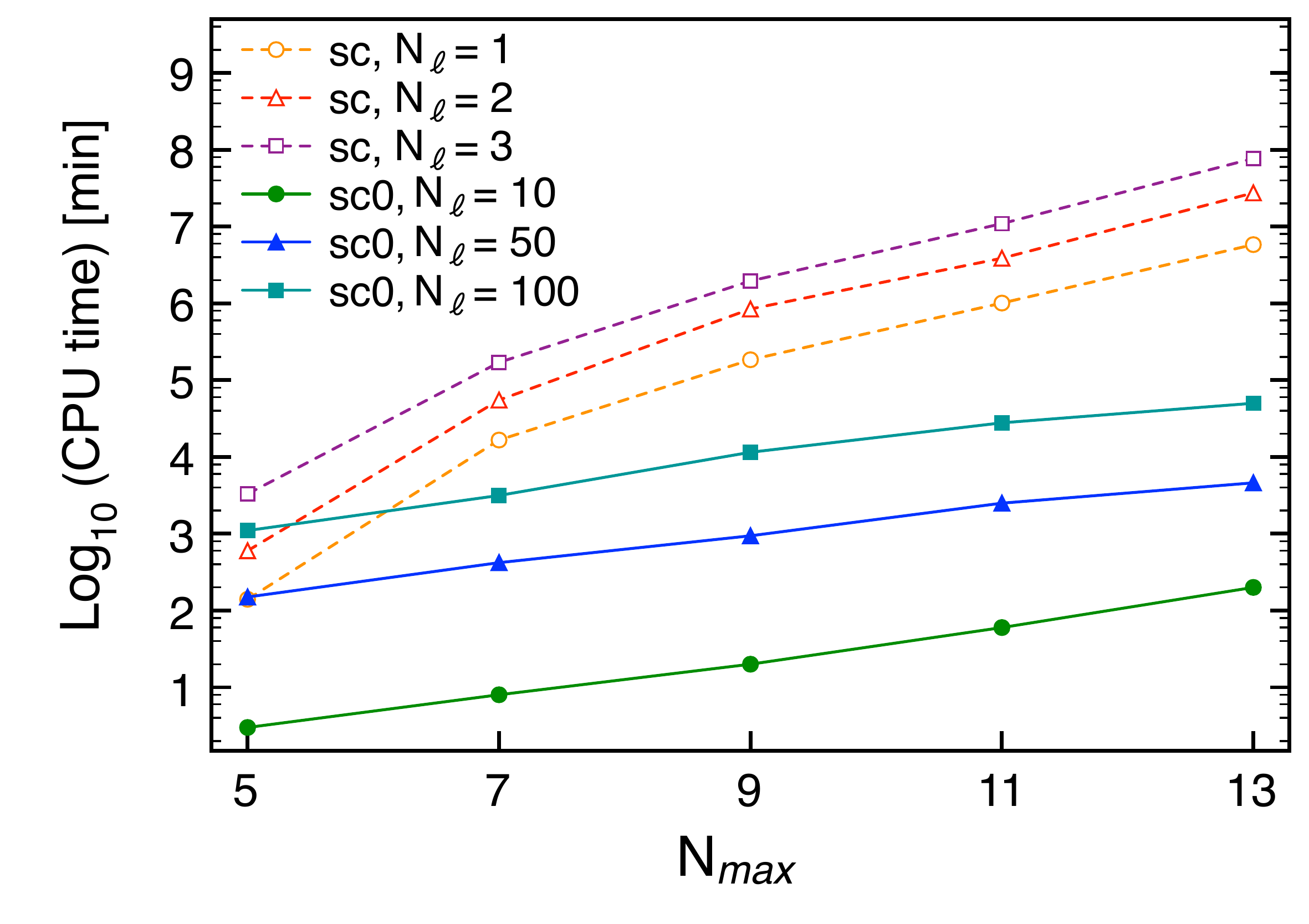}
	\caption{(Color online) CPU time requirements (in minutes) to perform 100 iterations within typical sc and sc0 self-consistency calculations. Results are shown for different numbers $N_\ell$ of Lanczos iterations per pivot as a function of the model-space size $\nmax$.}
\label{fig:cpu}
\end{center}
\end{figure}
The full sc scheme is significantly more expensive than the sc0 implementation. This is illustrated as a function of the model-space size $\nmax$ in Fig. \ref{fig:cpu} for typical Gorkov calculations performed using different numbers $N_\ell$ of Lanczos iterations per pivot. 
In fact, even when projecting the initial matrix onto a small Krylov space, the time required to run the sc scheme can easily become prohibitive in practice.
On the other hand, as discussed in Sec.~\ref{sec:self}, the sc0 scheme already grasps the relevant physics leading to accurate results (see also Refs.~\cite{Soma:2013rc,Cipollone:2013b}) and could be applied to larger model spaces necessary to handle SRG-unevolved NN interactions and/or heavy systems.
The sc0 scheme therefore constitutes an optimal choice for practical applications.


\section{Performance}
\label{sec:Performance}

As already mentioned, we assume the nuclei under study to be in a $J^\Pi = 0^+$ state and expand Gorkov's propagators on a spherical harmonic oscillator basis characterized by quantum number $a~= (n, q, j, m, \pi)\equiv(n, \alpha, m)$, where $n$ and $m$ label the principal quantum number and the projection of the third component of the angular momentum, respectively. As a result, Gorkov's equation can be written in a block diagonal form that separates out the partial waves $\alpha$.
Unless otherwise stated, the two-body potential employed is a next-to-next-to-next-to-leading-order (N$^{3}$LO) 2N chiral interaction~\cite{N3LO:2003,MacEnt2011}  ($\Lambda_{\chi}=$500\,MeV) complemented by the Coulomb force. 
The resulting isospin-symmetry breaking interaction is then softened using free-space similarity renormalization group (SRG) techniques~\cite{Bogner:2009bt} down to a momentum scale of \hbox{$\lambda = 2.0$ fm$^{-1}$}.

\subsection{Krylov projection}
\label{sec:lanczos}


\subsubsection{Choice of Lanczos pivots}
\label{subsec:pivots}

In exact arithmetic, the Lanczos algorithm generates basis vectors that are all orthogonal to each other, until the full original space is spanned. On a computer, the finite precision of the machine will at some point spoil the orthogonality, resulting in a set of linearly dependent vectors. This can be usually corrected, e.g., by means of selective orthogonalization techniques~\cite{Parlett:1979la}.
In the following we take instead a pedantic approach and orthogonalize after each iteration the new Lanczos vector with respect to all previous ones. This procedure is increasingly costly in the limit of large $N_L$, but it is doable and provides the safest option for actual calculations where one is interested in relatively small Krylov spaces.
An additional mechanism causing the sudden loss of orthogonality relates to the convergence of the eigenvalues of the Krylov-projected matrix, known as Ritz (eigen)values, to machine precision. This however happens only for extremely large spaces that approach the dimension of the original space [Eq.~\eqref{eq:limitnl} below] and does not affect in practice our Gorkov calculations. Nevertheless, we still check for sudden losses of orthogonality between successive Lanczos vectors\footnote{Following Ref.~\cite{Parlett:1979la}, two vectors $\mathbf{v}$ and $\mathbf{w}$ are considered orthogonal if $\mathbf{v} \cdot \mathbf{w} < \sqrt{\epsilon_m}$, where $\epsilon_m$ is the machine precision. In the case of the present calculations $\epsilon_m=1.11 \cdot 10^{-16}$.}.  If this occurs we stop the projection just before, at a corresponding number $N_L^{crit}$ of Lanczos iterations.

%
%

\begin{table}[t]
\centering
\begin{tabular}{|c|c|c|c|}
\hline
$N_\ell$ & ~$N_p$~ & $N_L$/$N_L^{crit}$ &  E$_{\nu s_{1/2}}$ [MeV]  \\
\hline
\hline
1  &  4 & 4 & -2.286045527516 \\
5  &  4 & 20 & -2.285370055029 \\
10  &  4 & 40 & -2.285503728538 \\
50 &  4 & 200 & -2.285578135207 \\
100  &  4 & 400 & -2.285580911804 \\
150  &  4 & 600 & -2.285580911686 \\
200  & 4  &  800 &   -2.285580911686 \\
297  &  4  & 1111$^*$ & -2.285580911686 \\
300  &  3 &  1121$^*$ & -2.285580911686 \\
400  &  3  & 1113$^*$ & -2.285558373049 \\
800  &  2 &  1103$^*$ &  -2.285504651650 \\
1000  &  2 & 1029$^*$ & -2.285580911687\\
1188  &  1 & 1029$^*$ & $\,\,\,$ -2.215766990937 $\,\,\,$ \\
\hline
\hline
\multicolumn{3}{|l}{Exact diagonalization:~~~} & -2.285580911686 \\
\hline
\end{tabular}
\caption{Contribution to the total binding energy from the neutron $s_{1/2}$ partial wave in $^{12}$C, for different numbers of
pivots used ($N_p$) and number of iterations per pivot ($N_\ell$). 
Asterisks ($^*$) indicate a truncation of the Lanczos iterations at $N_L^{crit}$ due to a sudden loss of orthogonality. Otherwise,
a total number $N_L$=$N_p\times N_\ell$ vectors is generated. The  dimension of the full 3QP space is $N_s^{\nu s_{1/2}}$=1188. }
\label{tab:loss1}
\end{table} 
\begin{table}
\centering
\begin{tabular}{|c|c|c|c|}
\hline
$N_\ell$ & ~$N_p$~ & $N_L$ &  E$_{\nu s_{1/2}}$ [MeV]  \\
\hline
 \hline
~600~  &  1 & 600 &             -2.109743018672  \\
  300    &  2 & 600 &             -2.268918978484 \\
  200    &  3 & 600 &             -2.279490387096 \\
  150    &  4 & 600 &  $\,\,\,$ -2.285580911686  $\,\,\,$\\
\hline
\hline
\multicolumn{3}{|l}{Exact diagonalization:~~~} & -2.285580911686 \\
\hline
\end{tabular}
\caption{Same as Tab.~\ref{tab:loss2} but for a fixed total number of Lanczos vectors and varying the number of linearly independent pivots.}
\label{tab:loss2}
\vspace{-.5cm}
\end{table}


A first basic test concerns the limit
\begin{equation}
\label{eq:limitnl}
\mbox{dim}(\mathcal{K}) \longrightarrow \mbox{dim}( E ) \:
\end{equation}
[see also Eq.~\eqref{eq:limitK}], where the Krylov subspace coincides with the initial one and the exact result must be recovered.
To this extent, we calculate the partial contribution of one specific channel to the binding energy of $^{12}$C, Eq.~\ref{eq:koltun_gorkov},
in a small model space where the Krylov projection can be compared to the exact diagonalization of the original matrix. Tables~\ref{tab:loss1} and~\ref{tab:loss2} list the contribution of neutron orbits characterized by $j^\pi=1/2^+$ ($\alpha$=$\nu s_{1/2}$) to the Koltun sum rule in a small model space of 4 major oscillator shells ($\nmax=3$) and for different numbers of iterations and pivots used.
In this case $N_s^{\nu s_{1/2}}=1188$, $N_b^{\nu s_{1/2}}=2$ and the total dimension of the HFB space is 4. Thus, only up to $N_p^{\nu s_{1/2}}=4$ Lanczos pivots can be generated from linearly independent vectors in the HFB space [Eq.~(\ref{eq:pivots})].
As long as the number of iterations per pivot, $N_\ell$, is small enough to allow for all the 
 $2 N_b^{\nu s_{1/2}}$ pivots to be used, the Krylov-projected energy converges to the exact value in the limit of Eq.~\eqref{eq:limitnl}. 
Table~\ref{tab:loss1} shows that $N_L=600$, which corresponds to half of the original 3QP configurations, is enough to recover the exact diagonalization  to thirteen significant digits.  Even for $N_\ell$=50, results are converged to better than 10~eV.  However, when $N_\ell$ increases a smaller number of pivots is exploited before the full space is saturated. 
The accuracy gradually worsens as the number of pivots used decreases, although results close to the exact one are found down to two pivots. In principle, one single pivot should be sufficient to recover the exact diagonalization in the limit \eqref{eq:limitnl}.
In practice, however, no more than a few \% accuracy is achieved before the loss of orthogonality kicks in. Conversely, adding just a few extra iterations of a second pivot brings the calculated energy close to the exact result.  The dependence of the result on the number of pivots used is shown in Tab.~\ref{tab:loss2} for a fixed dimension of the Krylov space. This demonstrates that the best possible accuracy is obtained when all linearly independent pivots are iterated. We further found that including all pivots is important to quickly converge off diagonal matrix elements of the self-energies, Eqs.~\eqref{eq:leh_self}, in accordance with the finding of Ref.~\cite{Schirmer1989}. This dependence on the number of pivots relates to having enough degrees of freedom to span the original HFB space, which is particularly important when resolvent operators are involved in the projection, as it is the case in Green's function theory.

In general, any set of linearly independent vectors in the HFB space can be used to generate the pivots through Eq.\eqref{eq:pivots}.
In our calculations, the optimal choice consists of using the HFB eigenstates themselves, which were indeed employed in the above tests. Vectors in the harmonic oscillator basis as well as random basis vectors lead to a worse convergence in all cases considered. Calculations of different partial waves, nuclei, interactions or model spaces validate the above findings. Given this, the $2 N_b$ HFB eigenstates are used as pivots throughout the following.

\subsubsection{$N_\ell$ dependence}
\label{subsec:nl}

When going to the large model spaces necessary to converge calculations with realistic nuclear interactions, currently available computational resources set severe limits on the dimension of matrix $\Xi$. A crucial issue concerns how large should the Krylov subspace be in order to achieve a satisfactory accuracy in the solution of Gorkov's equation. We now examine the dependence of the results on the number of Lanczos iterations per pivot, $N_\ell$.  We first do so on the basis of a single partial wave, as already done in connection with Tab.~\ref{tab:loss1}. Then, we investigate the convergence for a single Gorkov iteration but involving all partial waves at once. Finally, we terminate with complete self-consistent sc0 calculations.

For a given model space, the dimensions of both the 3QP space, $N^{\alpha}_s$, and the single-particle basis, $N^{\alpha}_b$, depends on the partial wave $\alpha=(q,j,\pi)$. For a fixed number of Lanczos iterations $N_\ell$,  the fraction of the initial space spanned by the Krylov-projected matrix depends on $\alpha$ as well. In general \hbox{$N_L^\alpha \propto  N_b^\alpha$} so that partial waves with low angular momentum will be better reproduced on average, since for a given $\nmax$ truncation the number of harmonic oscillator orbits $N_b^\alpha$ decreases with increasing $j_\alpha$. This is actually desirable because low angular-momentum waves correspond to the most occupied orbits and give the largest contributions to the binding energy. To quantify the fraction of the initial 3QP configuration space spanned by the Krylov projection for a given partial wave $\alpha$, we introduce
\begin{equation}
K^\alpha \equiv 100 * \frac{\mbox{dim}(\mathcal{K}{}^\alpha)}{\mbox{dim}(E^\alpha)} = \frac{100* N_\ell * N_p^\alpha}{\mbox{dim}(E^\alpha)} \, .
\label{eq:K}
\end{equation}

\begin{figure}[t]
\includegraphics[width=1.0\linewidth]{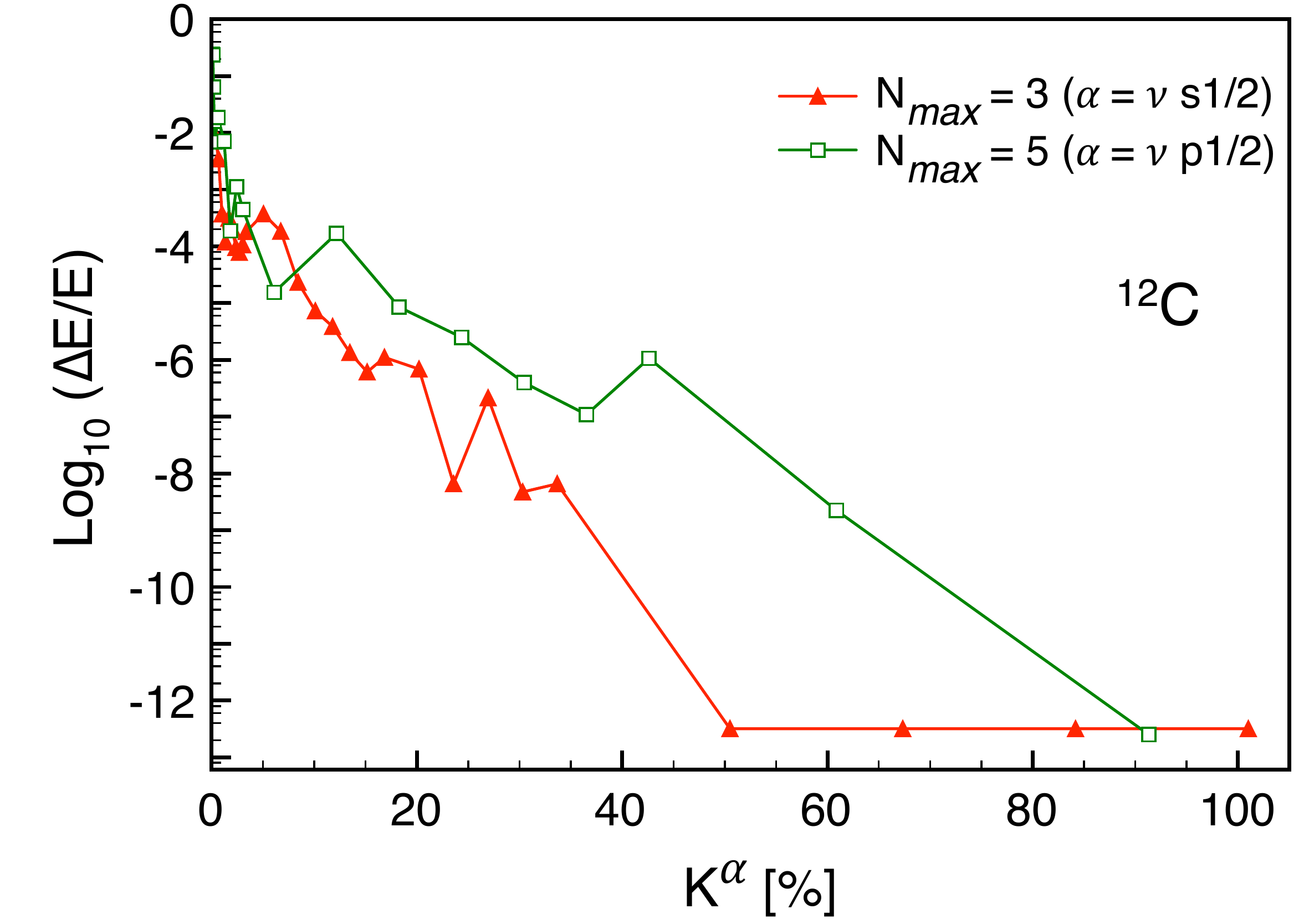}
\caption{(Color online) Relative error for the contribution of a given partial wave $\alpha$ to the Koltun sum rule in $^{12}$C
as a function of $K^\alpha$ (see text). Results refer to a single diagonalization and different model-space sizes. }
\label{fig:nlvsexact}
\end{figure}
Figure~\ref{fig:nlvsexact} displays in $^{12}$C the convergence of the contribution of two different partial waves to the Koltun sum rule, Eq~\eqref{eq:koltun_gorkov}, as a function of $K^\alpha$. Results are representative of how the error associated with a given partial-wave decreases by orders of magnitude when increasing $N_\ell$. Interestingly, relatively small values of $K^\alpha$ are sufficient to achieve precisions of the order of the keV in both cases. After this initial transient, the error follows a exponentially decreasing trend.
The $\nu s_{1/2}$ wave reaches the exact results up to machine precision when half of the 3QP space is projected to the Krylov subspace, as already seen in Table \ref{tab:loss1}. The convergence to the exact result is slower for the largest of the two model spaces used but the transient of the first few iterations remains.

To analyse the combined contributions from all partial waves, we now define the overall fraction of the 3QP space retained through
\begin{equation}
K'  \equiv 100 * \frac{\sum_{\alpha} \mbox{dim}(\mathcal{K}{}^\alpha)}{\sum_{\alpha} \mbox{dim}(E^\alpha)}
= \frac{100* N_\ell * \sum_{\alpha} N_p^{\alpha}}{\sum_{\alpha} \mbox{dim}(E^\alpha)} \, ,
\label{eq:Kprime}
\end{equation}
where $\alpha$ runs over all partial waves. Values obtained from  Eq.~\eqref{eq:Kprime} are displayed in Tab.~\ref{tab:dim} for different $\nmax$. For a fixed $N_\ell$, the fraction $K'$ becomes progressively small when increasing the size of the model space.
However, the total number of configurations still grows rapidly with $\nmax$.
\begin{table}
\begin{center}
\begin{tabular}{|c||c||c|c||c|}
\hline
$N_{\mbox{\footnotesize{max}}}$ & $\alpha_{\mbox{\footnotesize{tot}}}$
& $\sum_{\alpha}$  dim$(E^\alpha)$ &  $\sum_{\alpha} 2 N_b^{\alpha}$ & $K'(N_\ell=100) [\%]$ \\
\hline
\hline
3 & 7 & 12 226 & 20 & 16.358 \\
\hline
4 & 9 &  57 029 & 30 & 5.260 \\
\hline
5 & 11 &  411 968 & 42 & 1.019\\
\hline
7 & 15 &  3 265 512 & 72 & 0.220\\
\hline
9 & 19 &  16 808 456 & 110 & 0.065\\
\hline
11 & 23 & 65 305 228 & 156 & 0.023\\
\hline
13 & 27 &  \, 208 096 960 \, & 210 & 0.010\\
\hline
\end{tabular}
\end{center}
\caption{Values obtained from Eq. \eqref{eq:Kprime} for various model spaces. The sum over $\alpha$ is limited to neutrons only (including protons would require a factor 2 in columns 2, 3 and 4 that would cancel out in $K'$). As an example, $K'$ values for $N_\ell=100$ are displayed in the last column.}
\label{tab:dim}
\end{table}

\begin{figure}[t]
\includegraphics[width=1.0\linewidth]{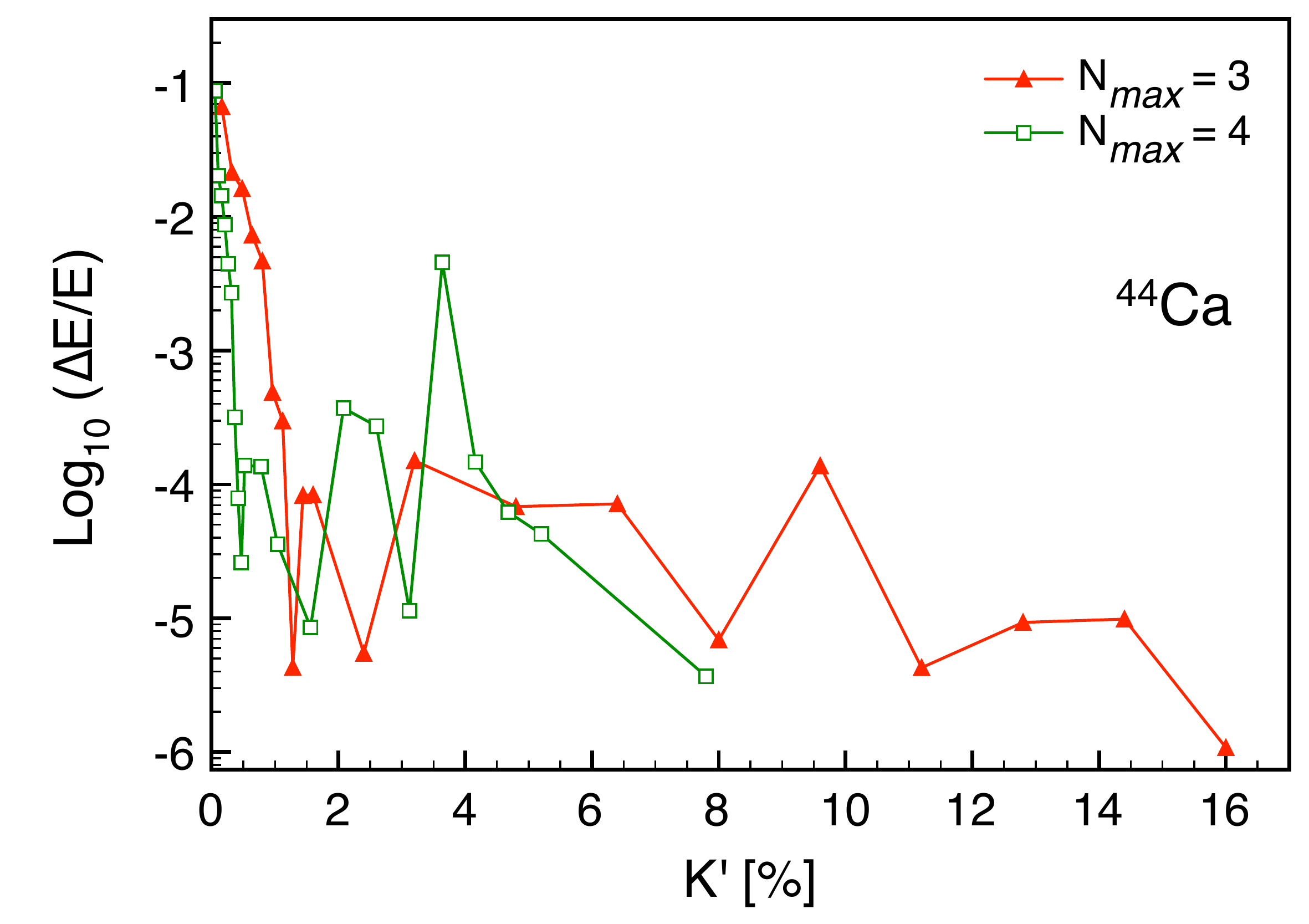}
\caption{(Color online) Relative error in the total binding energy of  $^{44}$Ca after one second-order iteration as a function of $K'$ (see text) for two different model-space sizes.}
\label{fig:allexact_scaled}
\end{figure}
\begin{figure}[h]
\includegraphics[width=1.0\linewidth]{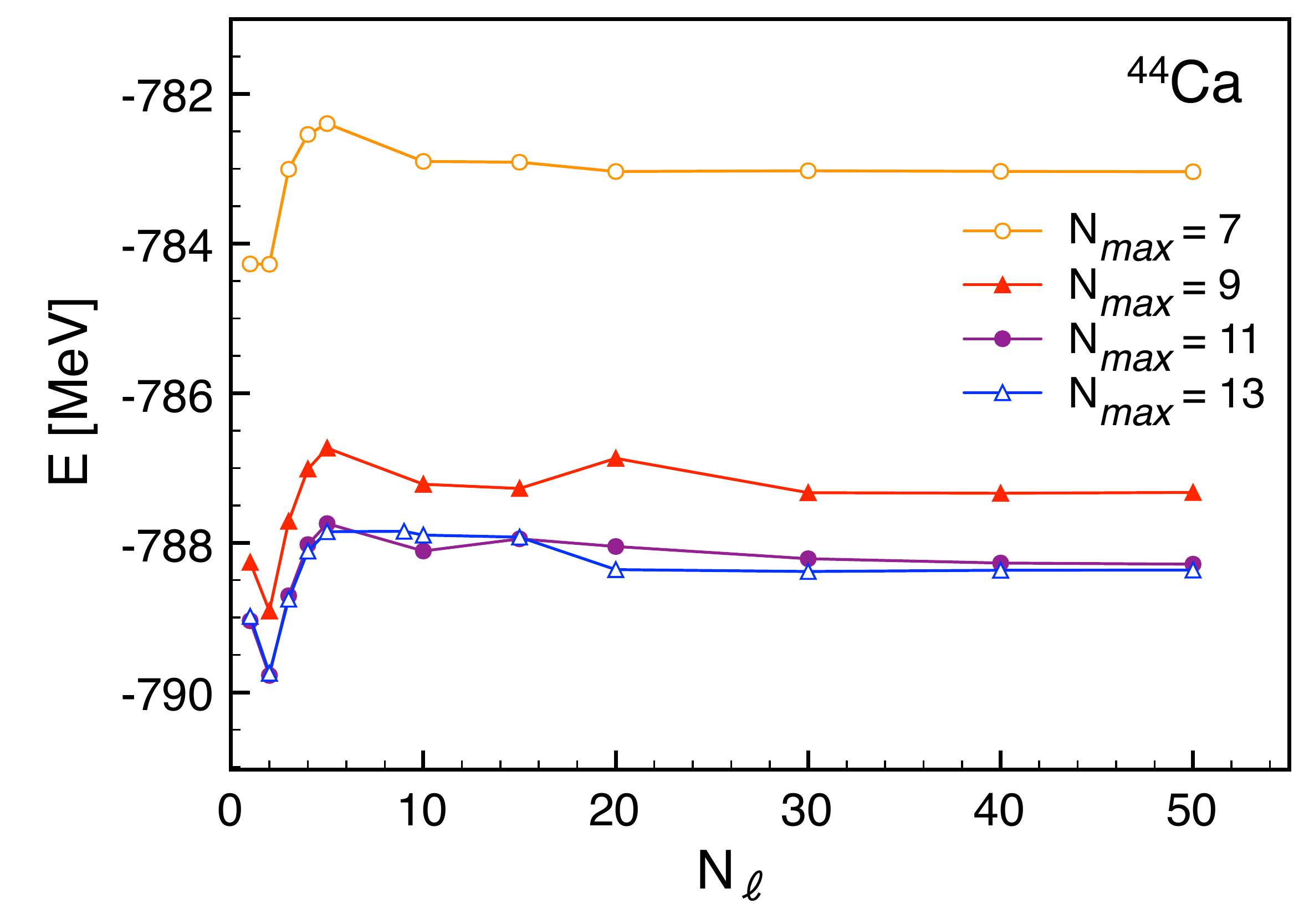}
\caption{(Color online) Convergence of the (sc0) binding energy of $^{44}$Ca as a function of $N_{\ell}$, for different model spaces.}
\label{fig:lanczos7-13}
\end{figure}
Figure~\ref{fig:allexact_scaled} demonstrates the accuracy obtained on the total binding energy as a function of $K'$, when all partial waves are accounted for in the calculation of $^{44}$Ca. 
Relative errors are given with respect to the result of one exact diagonalization in the original 3QP space. Errors for both $\nmax=3$ and $\nmax=4$ models spaces are comparable for $K'>\,$1\%  and eventually decrease in a similar fashion as in Fig.~\ref{fig:nlvsexact}. On the other hand, convergence to few keV is reached for smaller values of $K'$ in the larger model space.

Realistic calculations will differ from the above cases because diagonalizations have to be repeated iteratively to reach the self-consistent solution and because large model spaces must be employed.  In Fig.~\ref{fig:lanczos7-13}, converged sc0 energies are displayed as a function of $N_\ell$ for different model-space sizes. One notices that all cases show a similar dependence on $N_\ell$: a dip, a steep rise after $N_\ell=2$ and a smooth decay towards an asymptotic value. This behaviour is rather independent of $\nmax$ and indicates that $N_\ell$ is in fact a more appropriate parameter than $K'$ to gauge the convergence of the Krylov projection. Small fluctuations may still occur for $N_\ell > 10$, especially for the larger models spaces, which suggests that somewhat larger values of $N_\ell$ might be needed to reach the desired accuracy as $\nmax$ increases.
In general, this behaviour seen in Fig.~\ref{fig:lanczos7-13} is in accordance with the above observation that, when increasing $\nmax$, a smaller value of $K'$ is needed to reach a few keV accuracy. Arguably, binding energies are well reproduced once one includes the number of degrees of freedom sufficient to resolve the system's wave function (or propagator). The Krylov projection characterized by $N_\ell$ is a very efficient way to select those  degrees of freedom as it preserves the corresponding moments of the 3QP matrix $E$. 
The trend observed in Figs.~\ref{fig:nlvsexact} and~\ref{fig:allexact_scaled} suggest that $K'$ might instead control the exponential convergence to the exact diagonalization. From Fig.~\ref{fig:lanczos7-13} one sees that the energy reaches a plateau for $N_\ell > 30$, rather independently of the model-space size. Eventually, we estimate that the Lanczos procedure performed with $N_\ell \approx 50$ induces inaccuracies of about 100 keV for the largest model space considered ($\nmax=13$). 

\begin{figure}[t]
\begin{center}
        \includegraphics[width=1.0\linewidth]{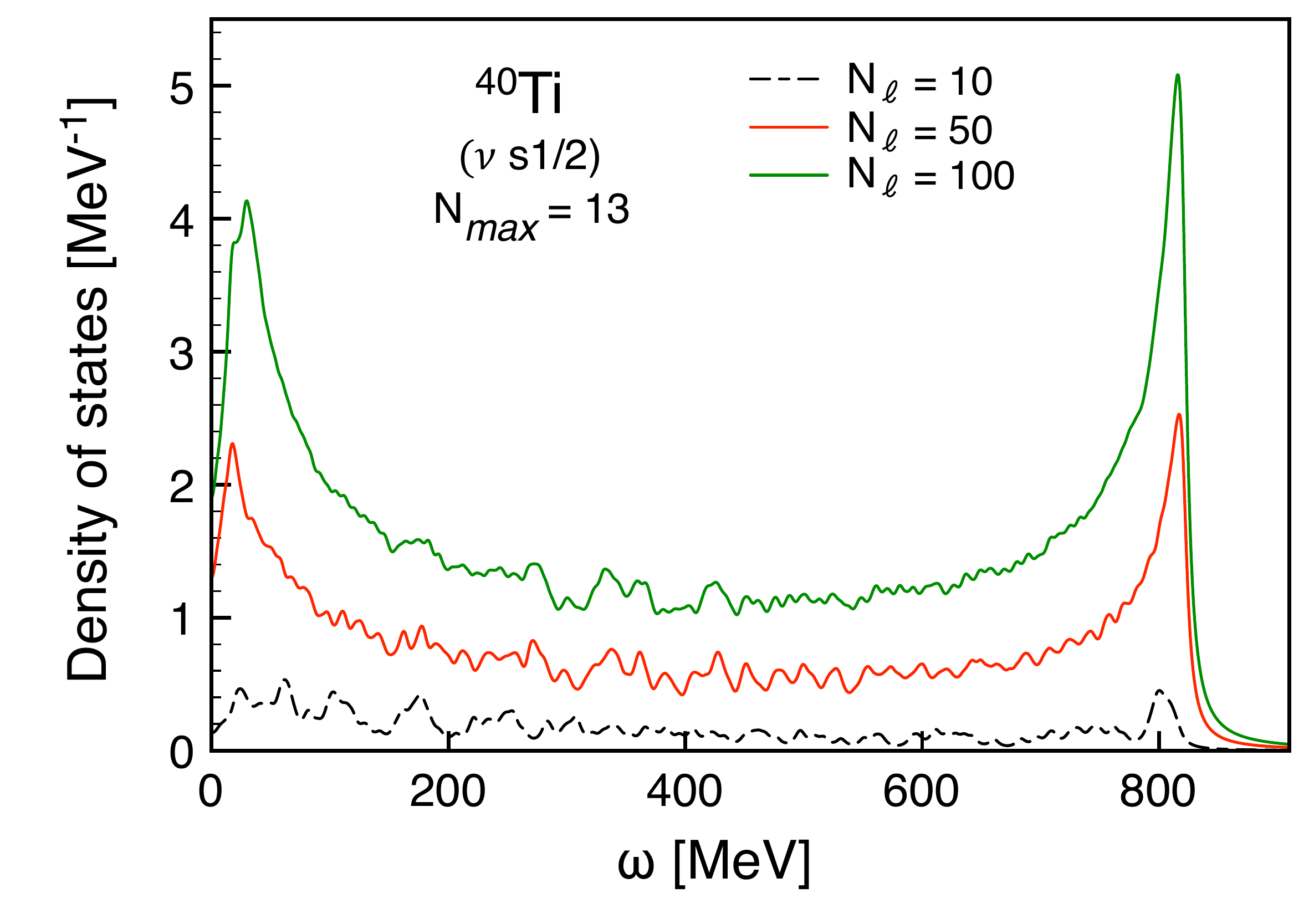}
	\caption{(Color online) Density of $J^{\Pi}=1/2^+$ states in $^{41}$Ti as a function of their excitation energy with respect to the Fermi level $\mu_n$ of  $^{40}$Ti, for increasing $N_\ell$. The distribution, discretized in the calculation, is convoluted with Lorentzian curves of 5~MeV width for display purposes.}
\label{fig:dos}
\end{center}
\end{figure}
\begin{figure}[h]
\begin{center}
        \includegraphics[width=1.0\linewidth]{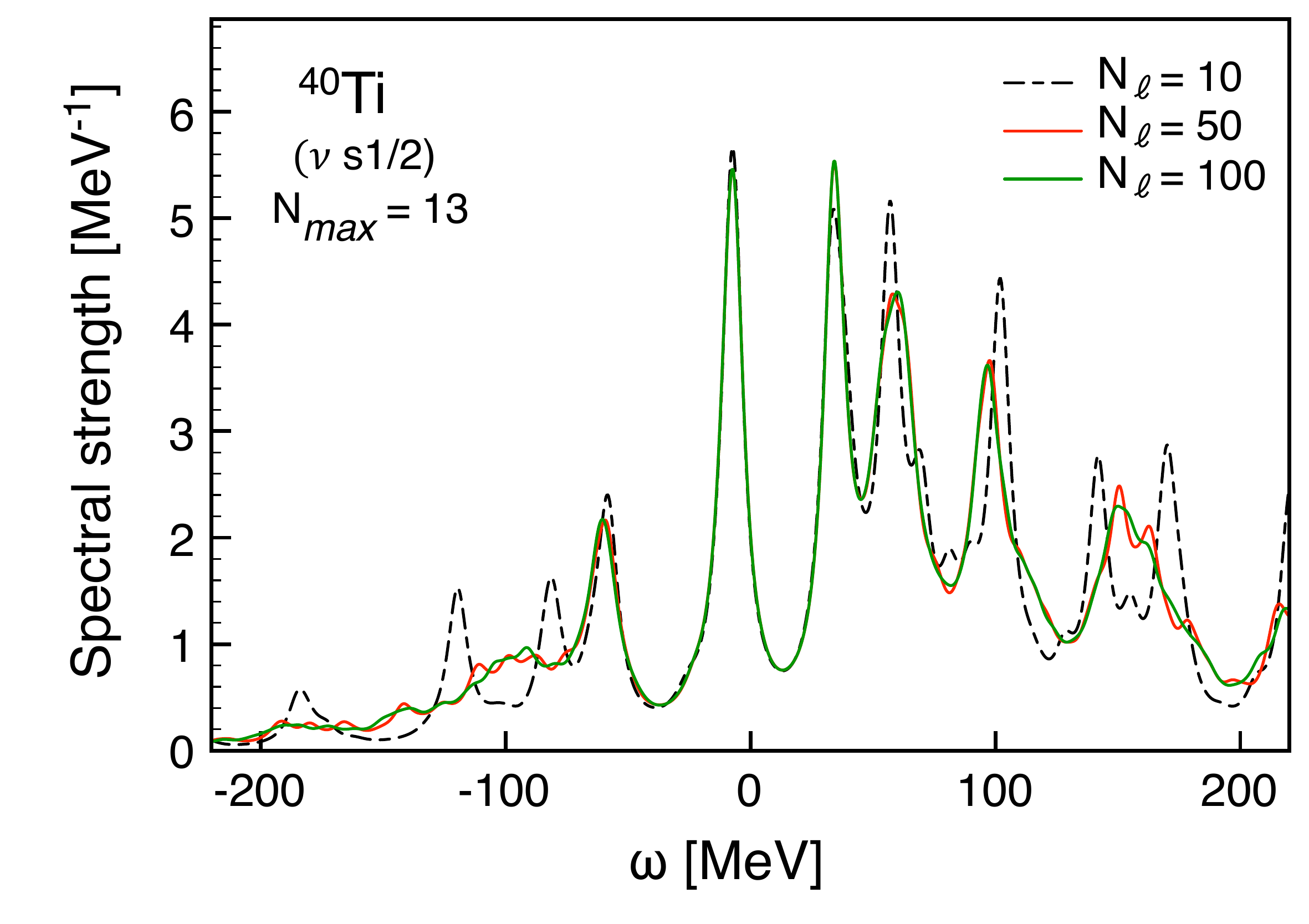}
	\caption{(Color online) One-neutron addition and removal spectral strength distribution in $^{40}$Ti limited to $J^{\Pi}=1/2^+$ final states in $^{39,41}$Ti. The distribution, discretized in the calculation, is convoluted with Lorentzian curves of 5~MeV width for display purposes.}
\label{fig:sp}
\end{center}
\end{figure}
It is also instructive to look at the convergence of spectroscopic quantities.  To this purpose, the doubly open-shell nucleus $^{40}$Ti is considered in a model space of 14 major shells. Figure~\ref{fig:dos} displays the density of $J^{\Pi}=1/2^+$ states\footnote{The density of states (DOS) in question is obtained from the SSD [Eq.~(\ref{eq:SSDan})] by setting $SF^{+}_k=1$ and $SF^{-}_k=0$ for all $k$.} in $^{41}$Ti as a function of their energy relative to the Fermi surface of $^{40}$Ti, for increasing $N_\ell$. The exact density of states would display a bell shape due to the rise of the number of (physical) degrees of freedom which is eventually stopped by the truncation of the model space. As seen from Tab.~\ref{tab:dim}, only a very small fraction of those configurations is effectively retained here. As the dimension of Gorkov-Krylov's matrix increases, only the density of states at the edges of the eigenvalue spectrum start to converge, which is a typical feature of Krylov methods. 

Despite the reduced DOS at the center of the spectrum, the spectral strength distribution [Eq.~(\ref{eq:SSDan})] is shown to converge rather rapidly at all energies when increasing $N_\ell$~\cite{CaurierRMP2005}. This is seen in Fig. \ref{fig:sp} where the neutron SSD in $^{40}$Ti, limited to $J^{\Pi}=1/2^+$ final states of $^{39,41}$Ti, is displayed. The curves obtained for $N_\ell=50$ and $N_\ell=100$ are essentially indistinguishable for most energies, with the SSD already converging to a resolution better than 10~MeV (5~MeV) for $N_\ell$=50 ($N_\ell=100$). Even for projections onto relatively small Krylov spaces, the result conserves the overall features of the SSD, which guarantees the quick convergence of observables and spectroscopic quantities in general.

\begin{figure}[t]
\begin{center}
        \includegraphics[width=1.0\linewidth]{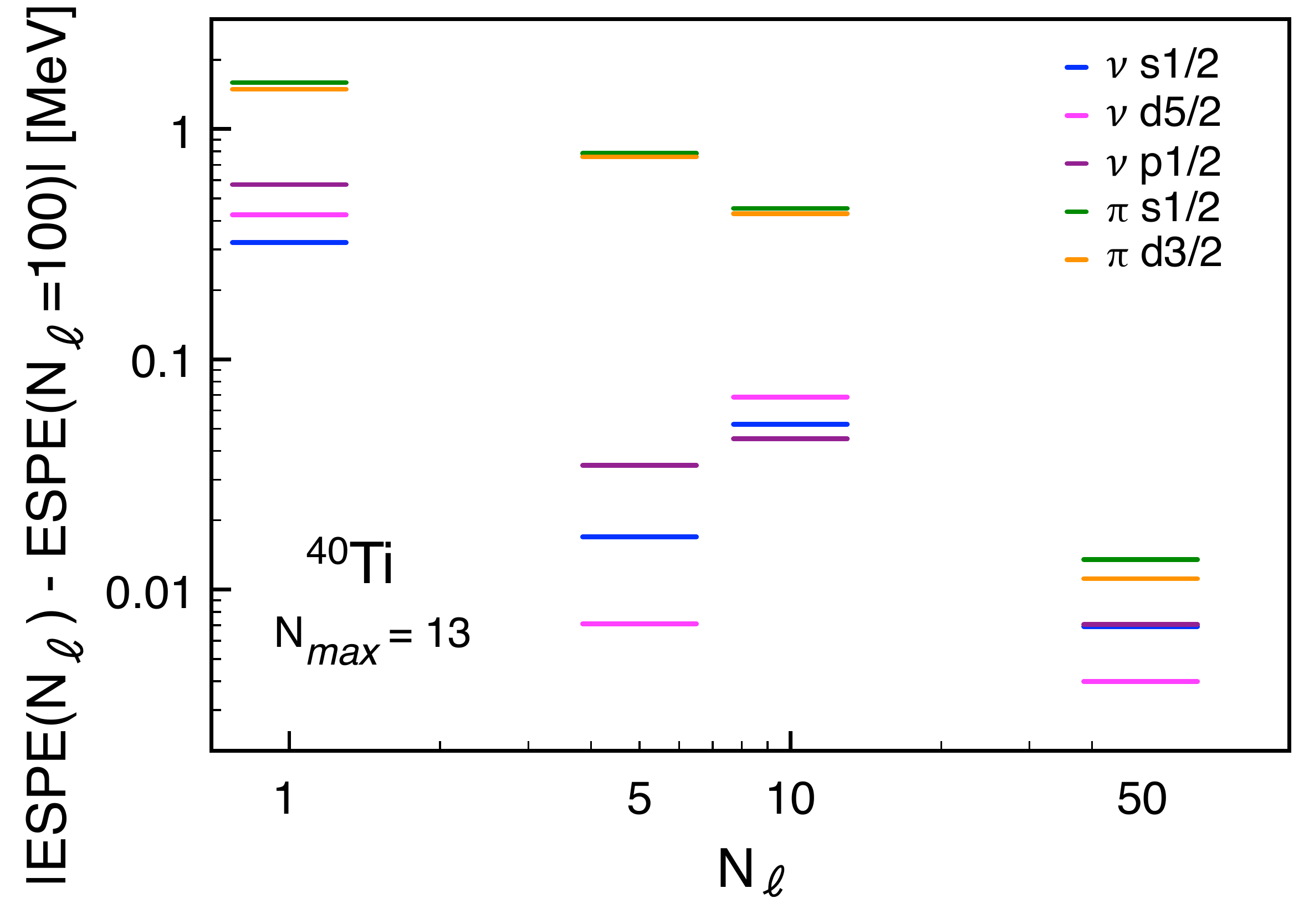}
	\caption{(Color online) Selected neutron and proton effective single-particle energies in $^{40}$Ti as a function of the number of Lanczos iterations per pivot $N_\ell$. Results are displayed relative relative to the values obtained for $N_\ell=100$. Calculations are performed in an $\nmax=13$ model space.}
\label{fig:espe}
\end{center}
\end{figure}
Figure~\ref{fig:espe} compares effective single-particle energies in $^{40}$Ti for different values of $N_\ell$. Results are given as the deviation to ESPEs computed for $N_\ell=100$, which is the most accurate truncation used. Difference between $N_\ell=10$ and $N_\ell=100$ are of the order of few tens to few hundreds keV, and decrease to 10~keV for $N_\ell$=50-100 for all ESPEs. This is also representative of the accuracy reached for one-nucleon separation energies
associated with the dominant quasiparticle states, which carry the main part of the strength.
Similar results are obtained for other nuclei and different model spaces.

Summarizing, the Krylov projection is shown to be reliable in all considered cases. The loss of orthogonality is well understood for small model spaces and never occurs in practice for large model spaces, where one is limited to a small number of Lanczos iterations. Both binding energies and one-nucleon separation energy spectra are well converged for relatively small values of $N_\ell$, nearly independently of the original dimension of Gorkov's matrix. This indicates that the Krylov projection is a reliable and computationally affordable tool that can be extended to large model spaces. For a typical large-scale calculation, a projection with $N_\ell=50$ is expected to yield a sufficient degree of accuracy for applications to mid-mass nuclei.
In this case, a conservative estimate of the systematic error induced by the projection is of the order of 300 keV on the converged total energy and 50 keV on one-nucleon separation energies associated with states carrying the dominant part of the strength, as well as on ESPEs. This can of course be improved by increasing~$N_\ell$.

\subsection{Self-consistency schemes}
\label{sec:self}

Section~\ref{subsec:calculation_scheme} outlines two different self-consistent calculation schemes. The sc implementation corresponds to a fully self-consistent solution of Gorkov's equation. Instead, the sc0 scheme iterates self-consistently only the static part of the self energy ${\bf \Sigma}^{(\infty)}$. A priori, there is no guarantee that one of these two many-body truncations will give results systematically closer to the exact binding energy than the other. However, the sc approach is conceptually superior both because it includes more diagrams (to very high orders) and because it guarantees that solutions satisfy fundamental conservation laws \cite{Baym:1961zz}.  From the computational cost point of view the sc0 approach is much more gentle than the sc scheme (see Fig.~\ref{fig:cpu}).

\begin{figure}[t]
\begin{center}
        \includegraphics[width=1.0\linewidth]{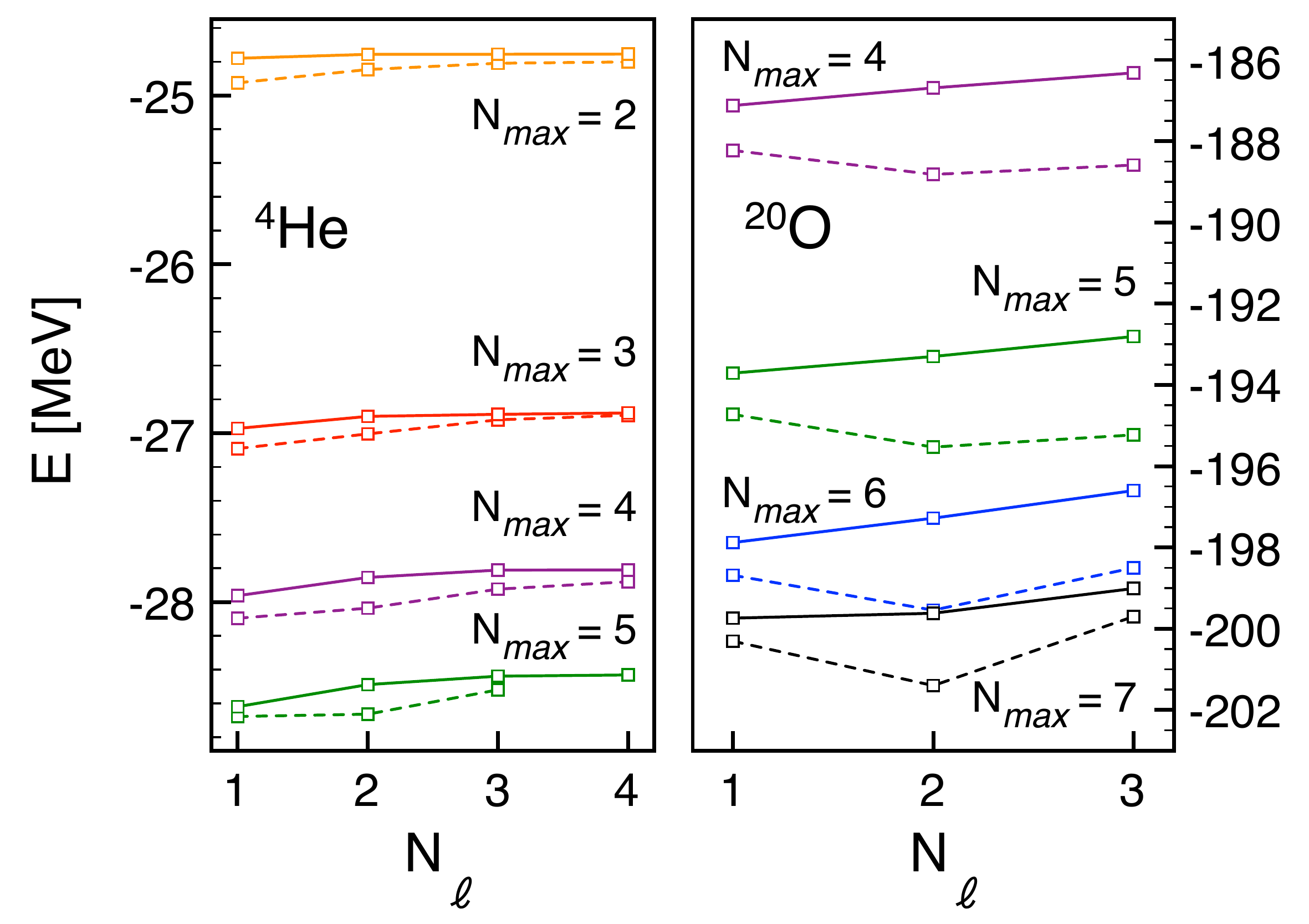}
	\caption{(Color online) Binding energy of $^4$He (left) and $^{20}$O (right) as a function of the number  $N_\ell$ of Lanczos iterations per pivot, for different model-space sizes. Dashed (solid) lines correspond to the sc (sc0) self-consistent scheme.}
\label{fig:sc-sc0}
\end{center}
\end{figure}
The two implementations are compared  in Fig.~\ref{fig:sc-sc0} for $^4$He and $^{20}$O. Results for total binding energies are displayed for different model spaces and small Krylov subspace projections, for which full sc calculations are possible. For all the cases considered, the two schemes yield results that differ at the level of 1$\%$. This is similar to the error induced by the many-body truncation employed in third- and higher-order SCGF calculations~\cite{Barbieri2012, Cipollone:2013b}. 
Thus, Fig.~\ref{fig:sc-sc0}  confirms the excellent performance of the partially self-consistent sc0 approach, making it an optimal compromise between high accuracy and an affordable computational cost.

\subsection{Model space convergence}
\label{sec:mod_sp}

The above discussion focused on the different technical steps that enable an efficient numerical solution of Gorkov's equation \eqref{eq:xi} for a given model space. We now turn to the convergence of Gorkov results as a function of the model space size.
\begin{figure}[t!]
\begin{center}
\includegraphics[width=1.0\linewidth]{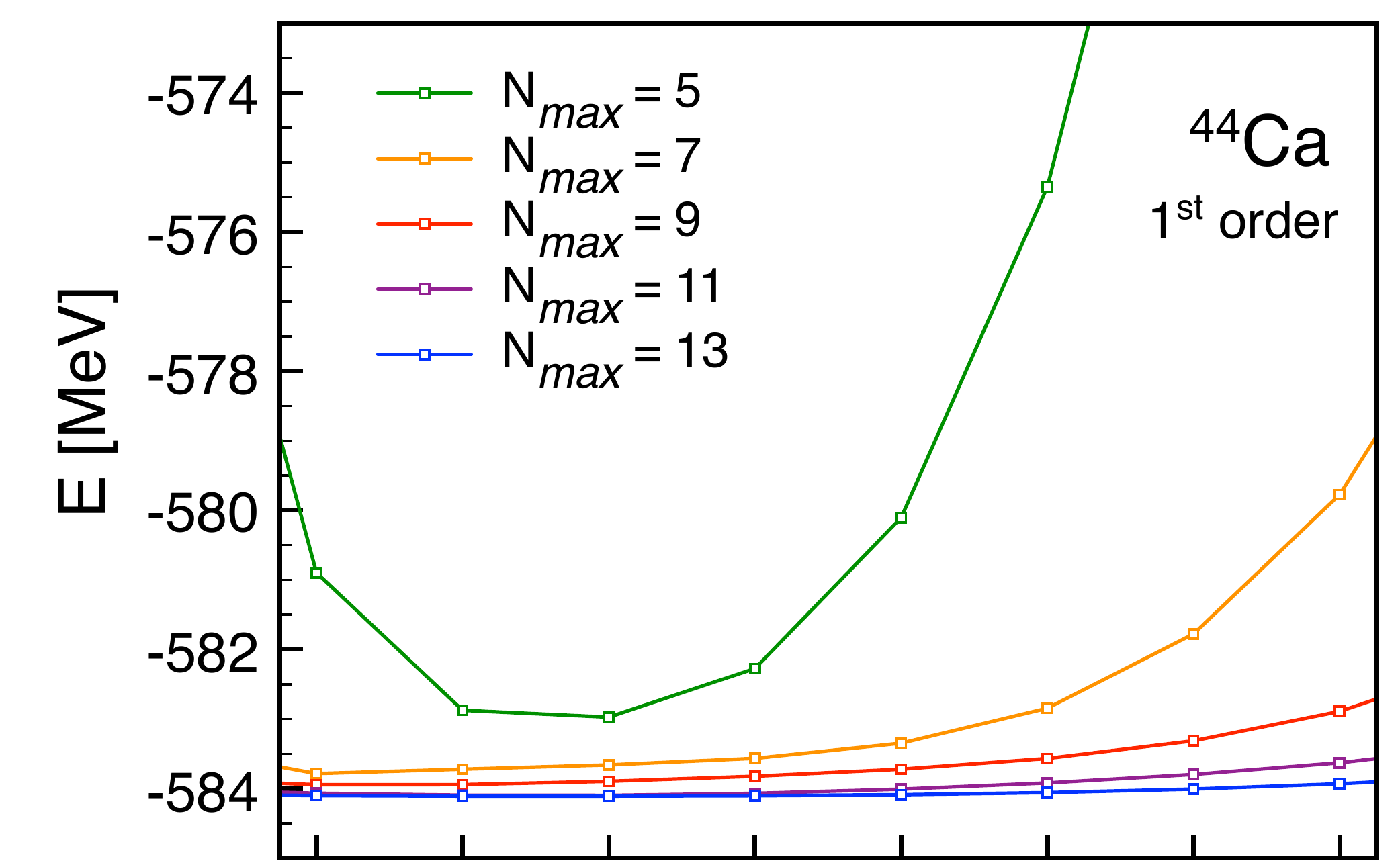}
\includegraphics[width=1.0\linewidth]{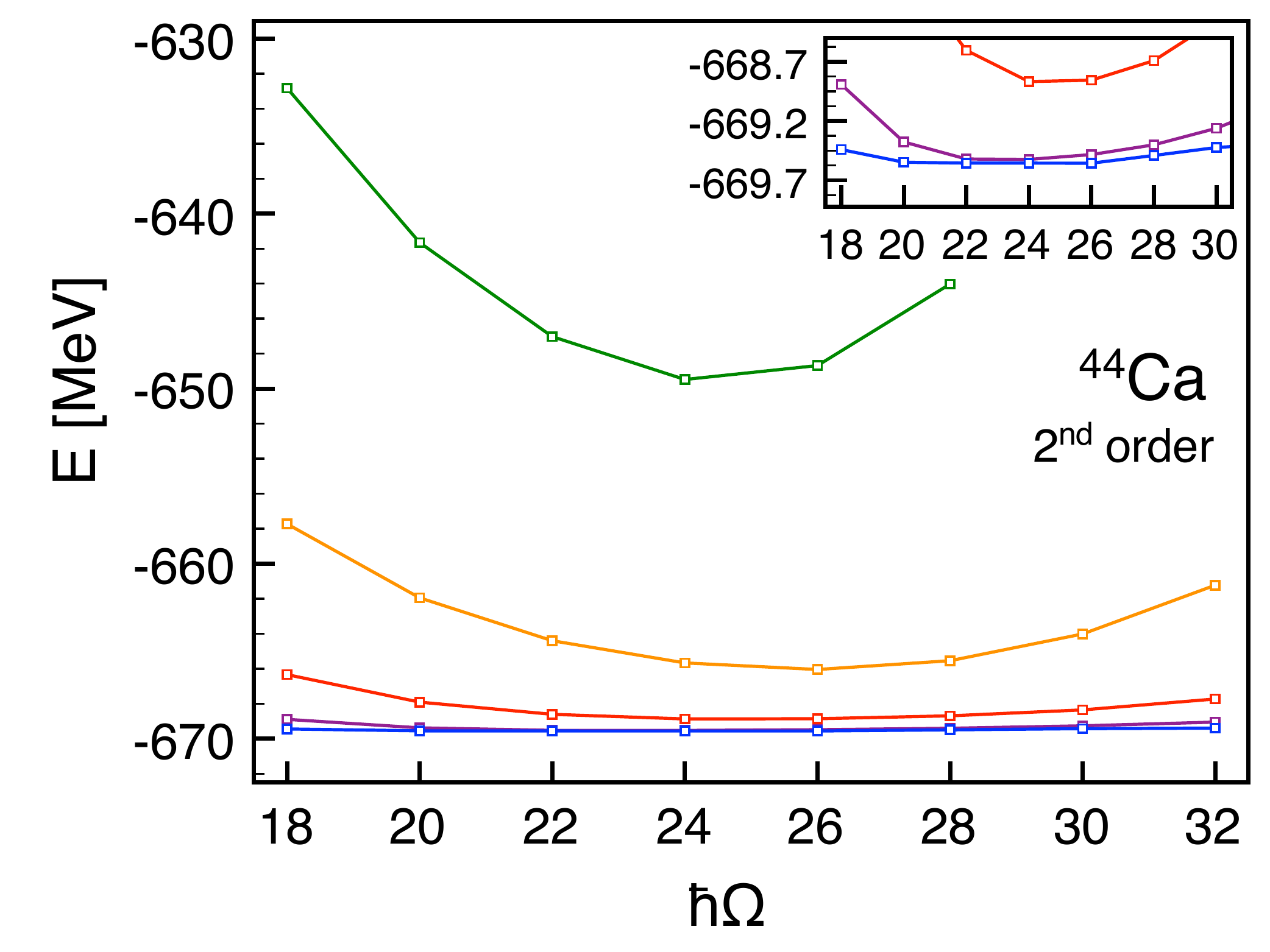}
\caption{(Color online) Binding energies of $^{44}$Ca from first-order (upper panel) and second-order (lower panel) Gorkov calculations as a function of the harmonic oscillator spacing $\hbar \Omega$ and for increasing size $\nmax$ of the single-particle model space. The insert shows a zoom on the most converged results.}
\label{fig:hwplots}
\end{center}
\end{figure}
\begin{figure}[h!]
\begin{center}
\includegraphics[width=1.0\linewidth]{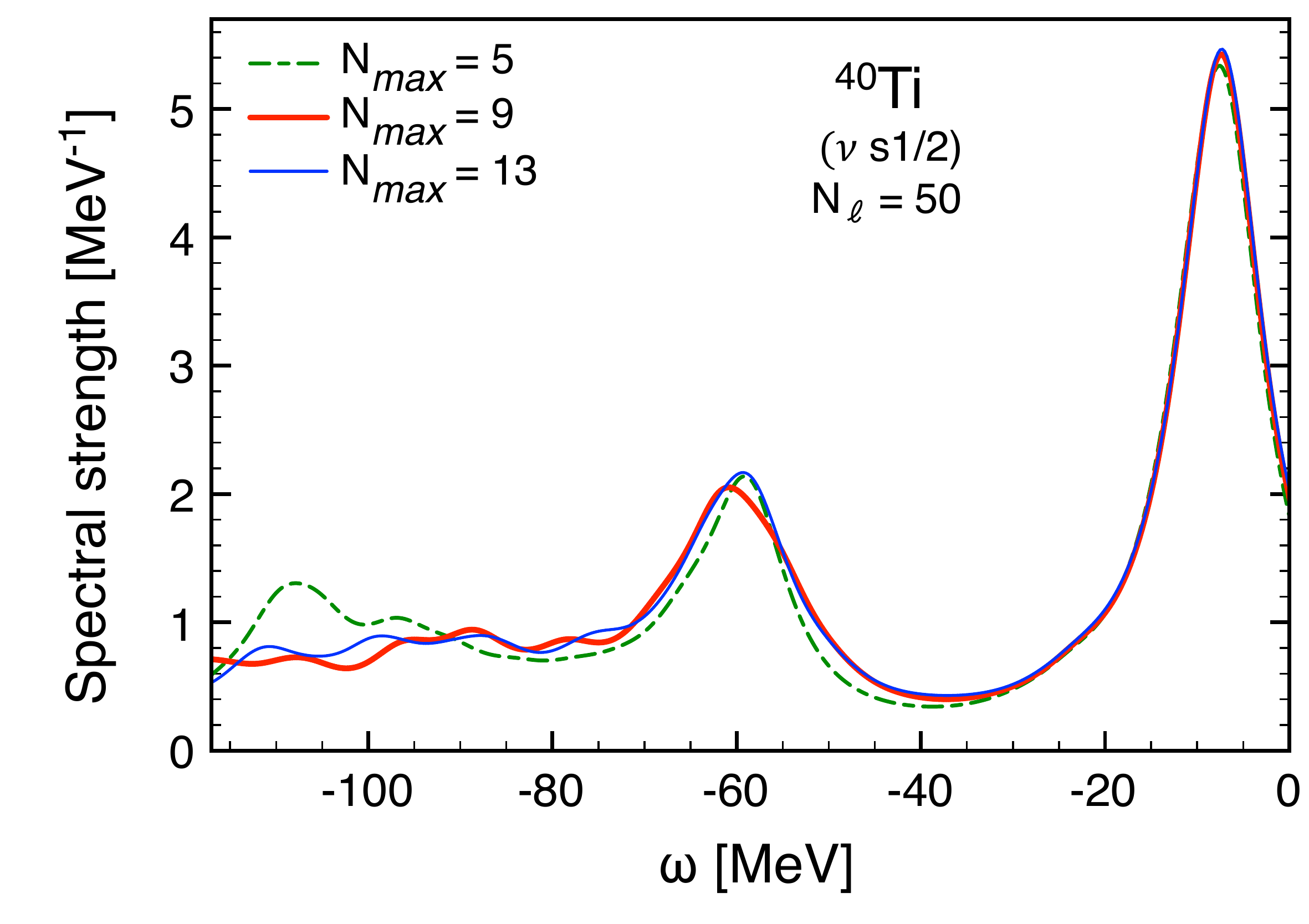}
	\caption{(Color online) $^{40}$Ti one-neutron removal spectral strength distribution associated with $J^{\Pi}=1/2^+$ final states as a function of the model space dimension $\nmax$. The distribution, discretized in the calculation, is convoluted with Lorentzian curves of 5~MeV width for display purposes.}
\label{fig:spnmax}
\end{center}
\end{figure}
For the harmonic oscillator model space considered here, this translates into requiring the independence of the results on the oscillator spacing $\hbar \Omega$ and  the number~$\nmax+1$ of major shells used. Figure~\ref{fig:hwplots} displays the convergence of the binding energy of the open-shell $^{44}$Ca nucleus. 
As $\nmax$ increases, results become independent of $\hbar \Omega$ and quickly converge to a fixed value. Energies generated at first (second) order in the self-energy expansion vary by only 10 keV (30 keV) when going from $\nmax=11$ to $\nmax=13$, well below the systematic error introduced by the Krylov projection. Similar convergence patterns have been found for closed-shell calcium isotopes, as well as for heavier systems such as $^{74}$Ni \cite{Soma:2013rc}.
Similar conclusions can be drawn for other quantities such as the SSD. In Fig.~\ref{fig:spnmax}, the neutron removal spectral strength distribution in $^{40}$Ti associated with $J^{\Pi}=1/2^+$ final states is plotted for different values of $\nmax$. Details of the SSD close to the Fermi surface ($\omega=0$) are well converged already at $\nmax=9$, while $\nmax=11$ is sufficient to converge the strength at very high (negative) energies to within the resolution set by the Lorentzian of 5\,MeV width used to convolute the spectrum.

For a given many-body method and truncation scheme, the convergence depends on the input NN (and 3N) interaction. In this sense, the robust behavior displayed in Fig. \ref{fig:hwplots} confirms the softness of SRG-evolved potentials used in this work, for which 14 major shells are sufficient to ensure well converged calculations. Our present implementation leaves room for improvement of the algorithms and better parallelisation so that the method can be pushed to larger model spaces. This presents opportunities for either going to even heavier systems or to employ interactions with a higher SRG cutoff. Both paths will be explored in forthcoming works.


\section{Conclusions}
\label{sec:conclusions}

We have detailed the numerical implementation of self-consistent Gorkov-Green's function (Gorkov SCGF) theory applied to finite nuclei~\cite{Soma:2011GkvI}. This  many-body method extends the reach of first-principle calculations to several hundreds of open-shell mid-mass nuclei that have been so far  inaccessible via \textit{ab initio} techniques. In this work, the numerical solution of Gorkov's equation of motion is discussed, paying particular attention to diagonalization, convergence and self-consistency issues. Numerical performances of this \textit{ab initio} scheme are analysed on the basis a N$^{3}$LO chiral nucleon-nucleon interaction, evolved down through similarity renormalization group (SRG) techniques to a resolution scale of \hbox{$\lambda = 2.0$ fm$^{-1}$}. Three-nucleon forces are currently being incorporated into Gorkov SCGF calculations following the recent step taken within Dyson's framework~\cite{Cipollone:2013b,Carbone:2013eqa}. We will report on this achievement in a forthcoming publication. While it is of importance to confirm the conclusions provided below when three-nucleon forces are in operation, the performance of the algorithm discussed in Secs.~\ref{sec:numerical} and~\ref{sec:lanczos} is of general character and should not be altered by the use of stronger two-nucleon interactions and/or of three-body forces.

A distinctive feature of Green's function methods is the automatic access to $\text{A}\pm1$ spectral strength distributions when computing the A-body ground-state. Self-energy contributions beyond first order account for dynamical correlations and induce the fragmentation of those spectral distributions. The self-consistent treatment of the fragmented strength requires a careful handling of the increasing number of poles generated at each iteration. Due to the doubling of the effective degrees of freedom associated with the use of Bogoliubov algebra, dealing with this issue within Gorkov's formalism is even more delicate than within Dyson's framework. The growth of the number of poles in Gorkov's propagators is controlled via Krylov projection techniques within a scheme that extends methods already in use in Quantum Chemistry~\cite{Schirmer1989}. The procedure can be executed to arbitrary accuracy, i.e. it recovers the exact result when the projection space coincides with the original one. The corresponding multi-pivot Lanczos algorithm is presented in details and its performances are tested thoroughly. 

The main conclusions reached in this work are that
\begin{enumerate}
\item Gorkov SCGF calculations performed on the basis of Krylov projection techniques display a favourable numerical scaling that authorizes systematic calculations of mid-mass nuclei. 

\item The corresponding multi-pivot Lanczos algorithm  is manageable from the numerical point of view, stable and was benchmarked favourably against  numerically exact solution of Gorkov's equation for small model spaces. The Krylov projection selects efficiently the appropriate degrees of freedom while spanning a very small fraction of the original space. For typical large-scale calculations of mid-mass nuclei, a Krylov projection employing $N_\ell\approx 50$  Lanczos iterations per pivot yields a sufficient degree of accuracy on binding energies, global features of spectral strength distributions, one-nucleon separation energies associated with states carrying the dominant part of the strength and on reconstructed effective single-particle energies. 

\item Fully self-consistent calculations of mid-mass systems in large enough model spaces are actually out of reach with currently available computational resources. A partially self-consistent scheme coined as "sc0" was designed and shown to reproduce well fully self-consistent solutions in small model spaces. The sc0 scheme provides an excellent compromise between accuracy and computational feasibility. Results of large-scale calculations shown in the present paper along with those already published~\cite{Soma:2013rc} or to be published in the future rely on this partially self-consistent scheme.  

\item The dependence of the results on the size of the harmonic oscillator model space was eventually investigated, showing a fast convergence for several observables of interest when employing SRG-evolved interactions. Given the gentle scaling of the numerical implementation we have developed, converged Gorkov SCGF calculations based on harder, e.g. original Chiral, interactions can be envisioned in the future.
\end{enumerate}

From the technical point of view, this work demonstrates that self-consistent Gorkov-Green's function calculations constitute a solid and viable candidate for the \textit{ab initio} description of medium-mass open-shell nuclei. The method has proven to perform well for both singly- and doubly-magic systems up to nickel isotopes~\cite{Soma:2013rc}.  The numerical scaling and performances offer the possibility to apply the method to even heavier systems in the future. Together with the on-going implementation of three-nucleon interactions and the development of a more accurate many-body truncation scheme, the present work sets the basis for systematic calculations of full isotopic and isotonic chains from an \textit{ab initio} perspective.

\acknowledgments

This work was supported by the United Kingdom Science and Technology Facilities Council (STFC) under Grants No. ST/I003363/1 and No.ST/J000051/1, by the DFG through Grant No. SFB 634, and by the Helmholtz Alliance Program, Contract No. HA216/EMMI. VS acknowledges support from Espace de Structure Nucl\'eaire Th\'eorique (ESNT) at CEA/Saclay. Calculations were performed using HPC resources from GENCI-CCRT (Grants No. 2012-050707 and 2013-050707) and the DiRAC Data Analytic system at the University of Cambridge (BIS National E-infrastructure capital grant No. ST/K001590/1 and STFC grants No. ST/H008861/1, ST/H00887X/1, and ST/K00333X/1).


\appendix

\section{Krylov projection}
\label{app:lanczos}

The critical step that allows for Gorkov calculations in large configuration spaces is the projection of the
3QP configurations space into a tractable Krylov subspace. Here, we present the details of the particular
Lanczos-based algorithm presently employed in Gorkov- and Dyson-Green's
functions calculations~\cite{Soma:2013rc,Barbieri:2009nx, Waldecker:2011by, Cipollone:2013b}.

When solving  Eq.~\eqref{eq:xi_12}, one needs to handle a matrix $E$ of large dimensions $N_s \times N_s$. 
%
%
Let $\mathcal{H}_{LG}$ be the space spanned by the eigenstates of $E$, with dim$(\mathcal{H}_{LG})=N_s$, and $\mathbf{p}$ 
a vector of dimension $N_s$ (usually referred to as the pivot). 
The Krylov subspace of order $r$ is the linear subspace of $\mathcal{H}_{LG}$ spanned by the images of $\mathbf{p}$
under the first $r$ powers of $E$, i.e.
\begin{equation}
\mathcal{K}^{(r)} \equiv \mbox{span} \left \{ \mathbf{p}, E \, \mathbf{p}, E^2 \, \mathbf{p}, E^3 \, \mathbf{p}, \dots ,
E^{r-1} \, \mathbf{p} \right \}\: .
\label{eq:krylov_1p}
\end{equation}
Provided that $E$ does not separates in sub-blocks of separate symmetry, one has that
\begin{equation}
\label{eq:limitK}
\mathcal{K}^{(N_s)} = \mathcal{H}_{LG} \: .
\end{equation}

The Lanczos algorithm is a procedure that generates an orthonormal basis $\{ \mathbf{v}_j; \, j=1,2, \dots r \}$
of $\mathcal{K}^{(r)}$ in the case where $E$ is Hermitian. Basis vectors $\mathbf{v}_j$
are obtained through a recursive procedure that involves vector-matrix multiplications, as follows
\begin{subequations}
\label{eq:Lanc_trad}
\begin{eqnarray}
\mathbf{v}_1 &\equiv& \mathbf{p}
\\
E \, \mathbf{v}_1 &\equiv& e_{11} \, \mathbf{v}_1 + e_{21} \, \mathbf{v}_2
\\
E \, \mathbf{v}_2 &\equiv& e_{12} \, \mathbf{v}_1 + e_{22} \, \mathbf{v}_2 + e_{32} \, \mathbf{v}_3
\\ \nonumber
&\dots&
\\
E \, \mathbf{v}_{r-1} &\equiv& e_{1(r-1)} \, \mathbf{v}_1 + \dots + e_{r (r-1)} \, \mathbf{v}_r \: \: ,
\end{eqnarray}
\end{subequations}
where at each step the newly generated vector $\mathbf{v}_j$ is further normalized to 1.
Following the above construction one has
\begin{equation}
e_{ij} = (e_{ji})^* =  \mathbf{v}_i^{\dagger} E  \, \mathbf{v}_j = 0 \qquad \mbox{for} \: \: |i-j| \geq 2 \: ,
\end{equation}
such that the projection $E'$ of the matrix $E$ on $\mathcal{K}^{(r)}$ is tridiagonal.

A similar procedure is applied here to reduce response operators such as Eq.~\eqref{eq:Krylov_prop}, 
where $E$ is defined in a large configuration space $\mathcal{H}_{LG}$ and 
the matrix product $\mathcal{C}\mathcal{C}^\dagger$ is defined in  a smaller space  $\mathcal{H}_{SM}$.
In this situation, it becomes necessary to exploit more than a single pivot vector to 
quickly  converge all degrees of freedom in $\mathcal{H}_{SM}$.
In our Gorkov calculations, $\mathcal{H}_{SM}$ is the HFB one-body Hilbert space,
which has twice the dimension of the single-particle basis employed. Thus,
we generate $N_p=2 N_b$ different vectors according to Eq.~\eqref{eq:pivots}. 
%
%

Let $\{\mathbf{p}^{(i)}; i=1, \dots N_p \}$ be a set of linearly independent vectors.
The new Krylov space is generated by extending the definition of Eq.~\eqref{eq:krylov_1p}
and the Lanczos procedure~\eqref{eq:Lanc_trad} to the case of multiple pivots.
Each vector $\mathbf{p}^{(i)}$  is thus iterated  a number
of times $r_i$, so that the total dimension of the basis generated is 
\begin{equation}
\label{eq:ri}
N_L = \sum_{i=1}^{N_p} \, r_i \:\: .
\end{equation}
In our algorithm, Lanczos iterations~\eqref{eq:Lanc_trad} are performed in sequence 
for each starting vector $\mathbf{p}^{(i)}$. It is therefore important that, at the starting of
each new set of iterations, the pivots are orthonormalized to the previously generated
basis vectors.

The first pivot $\mathbf{p}_1$ is simply iterated $r_1$ times as follow
\begin{subequations}
\label{eq:lan_mp}
\begin{eqnarray}
\mathbf{v}_1^{(1)} &\equiv& \mathbf{p}^{(1)}
\\
E \, \mathbf{v}_1^{(1)} &\equiv& e_{11} \, \mathbf{v}_1^{(1)} + e_{21} \, \mathbf{v}_2^{(1)}
\\ \nonumber
&\dots&
\\
E \, \mathbf{v}_{r_1}^{(1)} &\equiv& 
e_{(r_1 - 1) r_1} \, \mathbf{v}_{r_1-1}^{(1)}  + e_{r_1 r_1} \, \mathbf{v}_{r_1}^{(1)}
+ \mathbf{u}^{(1)} \: . \qquad
\end{eqnarray}
\end{subequations}
Up to this point the projected matrix $E'$ still maintains a tridiagonal structure and the vector 
$\mathbf{u}^{(1)}$ is orthogonal to the first $r_1$ basis vectors $\{\mathbf{v}^{(1)}_1, \ldots \mathbf{v}^{(1)}_{r_1}\}$.
As already mentioned,  $\mathbf{p}_2$ has first  to be orthogonalized with respect to the latters. Hence, one writes
\begin{equation}
\label{eq:piv2_ortho}
\mathbf{p}^{(2)} \equiv \sum_{i=1}^{r_1} c_i^{(1)} \, \mathbf{v}_i^{(1)} + d^{(1)} \, \mathbf{v}_1^{(2)} \: ,
\end{equation}
imposing $||\mathbf{v}_1^{(2)}||=1$, and takes $\mathbf{v}_1^{(2)}$ as the new pivot. Since 
$\mathbf{v}_1^{(2)}$ is orthogonal to all previous vectors, using the hermiticity of $H$ and
the tridiagonal form of Eqs.~\eqref{eq:lan_mp} one can prove that
\begin{subequations}
\label{eq:pivchng_12}
\begin{eqnarray}
\mathbf{v}_i^{(1) \, \dagger} E \, \mathbf{v}_1^{(2)} &=& 0 \qquad \forall \; i=1,\dots,r_1-1 \: ,
\\
\mathbf{v}_{r_1}^{(1) \, \dagger} E \, \mathbf{v}_1^{(2)} &=& \mathbf{u}^{(1) \dagger} \mathbf{v}_1^{(2)} =  e_{r_1 (r_1+1)} \: .
\end{eqnarray}
\end{subequations}
In general,  each vector $\mathbf{p}^{(i)}$, with $i \geq 2$, will be orthonormalised to the 
previously generated portion of the basis according to
\begin{equation}
\label{eq:pivAll_ortho}
\mathbf{p}^{(i)} \equiv \sum_{j=1}^{i-1} \sum_{k=1}^{r_{j}} c_k^{(j)} \, \mathbf{v}_k^{(j)} + d^{(i)} \, \mathbf{v}_1^{(i)} \: ,
\end{equation}
and the vector $||\mathbf{v}_1^{(i)}||=1$  is taken as the new pivot, which is iterated $r_i$ times.
If  $n_i$ is then number of basis vectors generated from all iterations before the $i^{\textrm th}$ pivot,
\begin{equation}
\label{eq:ri}
n_i = \sum_{j=1}^{i-1} \, r_j \:\: ,
\end{equation}
the iteration of pivot $\mathbf{v}_1^{(i)}$ yelds
\begin{subequations}
\begin{eqnarray}
E \, \mathbf{v}_1^{(i)} &\equiv&  \sum_{j=1}^{i-1} e_{(n_j + r_j) (n_i+1)  } \, \mathbf{v}_{r_j}^{(j)} \nonumber  \\
&& + e_{(n_i+1)(n_i+1) } \, \mathbf{v}_1^{(i)}  + \, e_{(n_i+2)(n_i+1) } \, \mathbf{v}_2^{(i)}
\qquad \qquad \\
E \, \mathbf{v}_2^{(i)} &\equiv&   \sum_{j=1}^{i-1} e_{(n_j + r_j) (n_i+2) } \, \mathbf{v}_{r_j}^{(j)} \nonumber \\
&& + e_{(n_i+1)(n_i+2)} \, \mathbf{v}_1^{(i)} \nonumber   \\
&& + \, e_{(n_i+2) (n_i+2)} \, \mathbf{v}_2^{(i)} + \, e_{(n_i+3) (n_i+2)} \, \mathbf{v}_3^{(i)}
\\ \nonumber
&\vdots&
\\
E \, \mathbf{v}_{r_i}^{(i)} &\equiv& \sum_{j=1}^{i-1} e_{(n_j + r_j) (n_i+r_i) } \, \mathbf{v}_{r_j}^{(j)} \nonumber  \\
&& 
+ e_{(n_i+r_i-1 ) (n_i+r_i)} \, \mathbf{v}_{r_i-1}^{(i)}+ e_{(n_i+r_i) (n_i+r_i)} \, \mathbf{v}_{r_i}^{(i)}  \nonumber \\
&&   + \, \mathbf{u}^{(i)} \: \: ,
\end{eqnarray}
\end{subequations}
where $\mathbf{u}^{(i)}$ is orthogonal to all previous vectors.

\begin{figure}
\vspace{-.9cm}
\begin{tabular}{m{0.7cm}m{6.2cm}m{0.55cm}}
\multirow{4}{*}{\vspace{-4.25cm}
$E' = $} &
\multirow{4}{*}{ \includegraphics[width=0.94\linewidth]{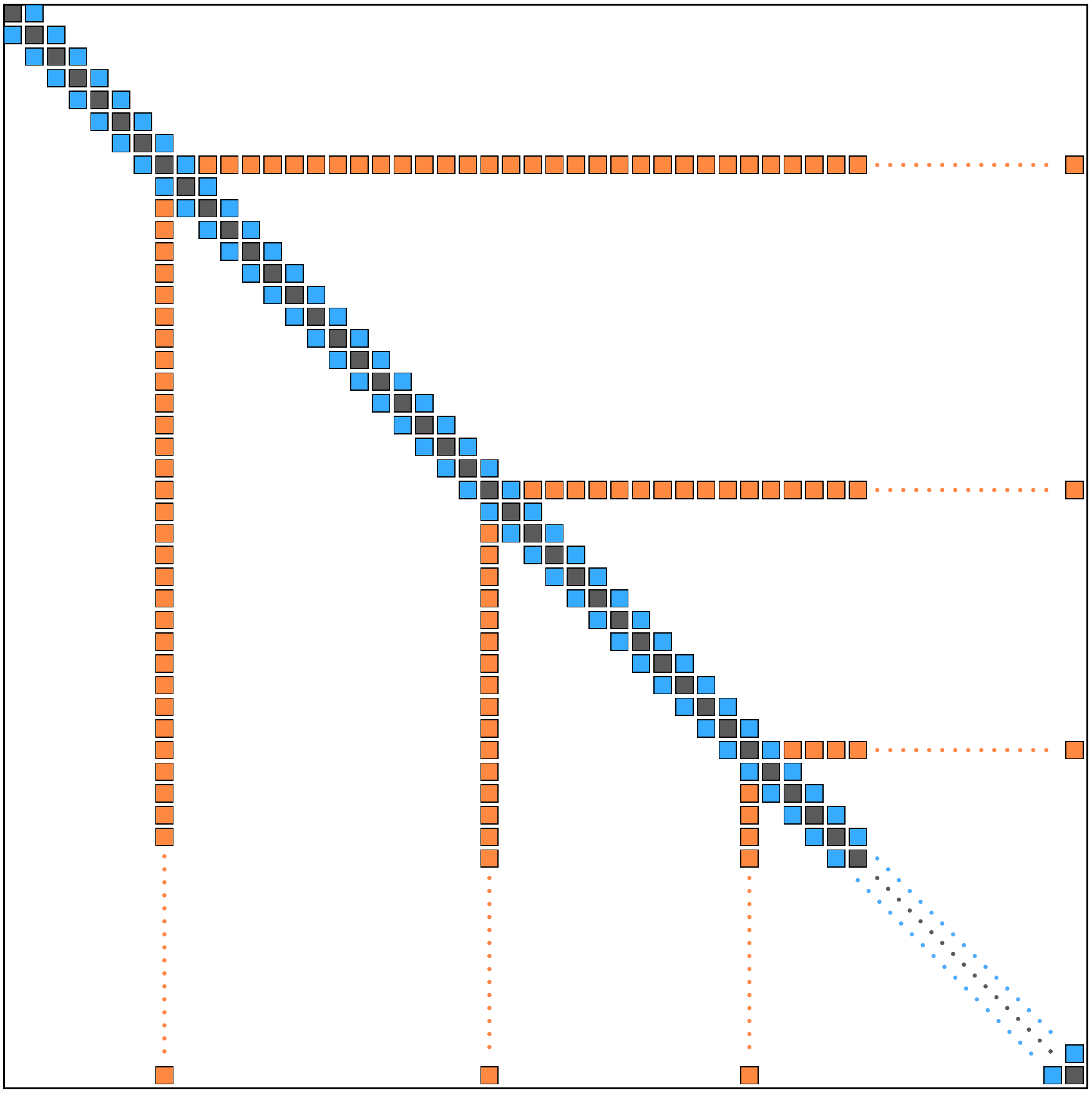}} &
\vspace{1.15cm}
$\hspace{-0.52cm} \left. \mbox{\fontsize{25}{10}\selectfont \phantom{I}} \right\} r_1$
\vspace{0.65cm}
\\
&& 
\vspace{-0.65cm}
$\hspace{-0.7cm} \left. \mbox{\fontsize{40}{10}\selectfont \phantom{I}} \right\} r_2$
\vspace{0.63cm}
\\
&& 
\vspace{-0.62cm}
$\hspace{-0.61cm} \left. \mbox{\fontsize{32}{10}\selectfont \phantom{I}} \right\} r_3$
\vspace{0.5cm}
\\
&& \vspace{0.5cm}
$\phantom{Z}$
\vspace{0.5cm}
\\
&
\multirow{4}{*}{\vspace{1.2cm} $\underbrace{\hspace{5.8cm}}_{\mbox{\small{$N_L$}}}$ \vspace{1.2cm} } 
\end{tabular}
\vspace{.2cm}
\caption{(Color online) Fishbone-like structure of the Lanczos reduced matrix $E'$.}
\label{fig:fishbone}
\end{figure}

Relations analogous to Eqs.~\eqref{eq:pivchng_12} holds every
time one moves to a new pivot, which connects the $\mathbf{v}_{r_i}^{(i)}$ vectors
(at the end of each block of iterations) to the remaining basis vectors. 
It follows that the tridiagonal form of the projected matrix $E'$ is maintained except
for the rows and columns corresponding to the last iteration of each pivot, which are nonzero
and give rise to the fishbone-like sparse matrix shown in Fig.~\ref{fig:fishbone}.

Notice also that the resulting space is not directly generated by the 
$\mathbf{p}^{(i)}$ vectors of Eq.~\eqref{eq:pivots} since these are 
othogonalized before they are iterated. 
Hence, the actual Krylov space is the one associated with the pivots
$\{\mathbf{v}_1^{(i)}; i=1, \ldots, N_p\}$
and it is defined as
\begin{eqnarray}
\label{eq:krylov_mult}
\nonumber
\mathcal{K}^{(r)} \equiv &\mbox{span} &\left \{ 
\mathbf{v}_1^{(1)}, E \, \mathbf{v}_1^{(1)}, E^2  \, \mathbf{v}_1^{(1)}, \dots , E^{r_1-1} \, \mathbf{v}_1^{(1)}  \, ,
\right.  \\
\nonumber &&\left . \phantom{ \{}
\mathbf{v}_1^{(2)}, E \, \mathbf{v}_1^{(2)}, E^2 \, \mathbf{v}_1^{(2)}, \dots , E^{r_2-1} \, \mathbf{v}_1^{(2)}  \, ,
\right.  \\
\nonumber &&\left . \phantom{\{ \{}
\dots
\right.  \\ 
& &\left. \phantom{ \{}
\mathbf{v}_1^{(N_p)}, E \, \mathbf{v}_1^{(N_p)}, \hspace{.3cm} \dots , \hspace{.3cm}  E^{r_{N_p}-1} \, \mathbf{v}_1^{(N_p)}
\right \} . \qquad \:\:
\end{eqnarray}
In the present work we choose a fixed number of iterations, i.e.  $r_i=N_\ell , \; \forall \, i=1, \ldots N_p$,
except for cases where a truncation of the Lanczos procedure required a lower number of iterations
for the last pivot (bottom part of Tab.~\ref{tab:loss1}).


\section{Adjustment of chemical potentials}
\label{app:mu}

Searching for the solution of Gorkov's equation, proton and neutron chemical potentials must be adjusted at each iteration in order to have the desired number of particles on average (see point 6 of the algorithm in Sec.~\ref{subsec:calculation_scheme}). After self-energies have been computed and Gorkov's matrix has been diagonalized, the average numbers of neutron and proton are evaluated through
\begin{subequations}
\begin{eqnarray}
\label{eq:NV}
N^{av} &= \displaystyle \sum_a^{\mbox{\tiny{neutrons}}} \, \rho_{aa} 
&= \sum_{a,k}^{\mbox{\tiny{neutrons}}} \left| {\mV_{a}^k} \right|^2 \: , \\
Z^{av} &= \displaystyle \sum_a^{\mbox{\tiny{protons}}} \, \rho_{aa} 
&= \sum_{a,k}^{\mbox{\tiny{protons}}} \left| {\mV_{a}^k} \right|^2 \: .
\end{eqnarray}
\end{subequations}
The resulting numbers are compared to the expected $N$ and $Z$. Chemical potentials $\mu_N$ and $\mu_Z$ are then increased (decreased) if the computed number of particle is smaller (larger) than the required values according to
\begin{equation}
\label{eq:adjust_mu}
\mu^{new}_{N,Z} = \mu^{old}_{N,Z} + \Delta^{\mu}_{N,Z} \: ,
\end{equation}
where
\begin{subequations}
\begin{eqnarray}
\label{eq:adjust_mu_delta}
\Delta^{\mu}_{N} &\equiv& \displaystyle C^{\mu}_{N} \, \frac{N - N^{av}}{N}  \: , \\
\Delta^{\mu}_{Z} &\equiv& \displaystyle C^{\mu}_{Z} \, \frac{Z - Z^{av}}{Z}  \: .
\end{eqnarray}
\end{subequations}
Parameters $C^{\mu}_{N,Z}$  control the speed and pattern of convergence, and are typically of order of unity. As long as the convergence is reached, the choice of $C^{\mu}_{N,Z}$ does not impact the final result.

Notice that subsequent adjustments of the chemical potentials may be necessary before the required precision of $N^{av},Z^{av}$ is achieved, implying that the above procedure is repeated several times at each self-consistent iteration. However, one is not interested (at least in the first few iterations) in having extremely precise neutron and proton numbers as the self-consistency process will make the optimal chemical potentials vary until a sufficient degree of self-consistency is reached.


\bibliography{./gorkovbib}

\end{document}